\documentclass[twocolumn,10pt]{IEEEtran}
\usepackage{amsmath,amssymb,amsthm}
\ifdefined\XeTeXversion
  \usepackage{fontspec}
  \IfFontExistsTF{Times New Roman}{\setmainfont{Times New Roman}}{}
\fi
\usepackage{graphicx}
\usepackage{booktabs}
\usepackage{cite}
\usepackage{url}
\usepackage{placeins}
\usepackage[hidelinks]{hyperref}
\hypersetup{
  pdftitle={Quantum Shock and Train-Time Constrained Quantization: A Geometric Study of Finite-Resolution Variational Quantum Circuits},
}

\usepackage{cleveref}


\newtheorem{theorem}{Theorem}
\newtheorem{corollary}{Corollary}
\newtheorem{proposition}{Proposition}
\newtheorem{lemma}{Lemma}

\newcommand{\ket}[1]{\left|#1\right\rangle}

\newcommand{\bm}[1]{\vec{#1}}
\newcommand{\dfs}{d_{\mathrm{FS}}}
\newcommand{\abures}{A_{\mathrm B}}
\newcommand{\gammab}{\gamma_{\mathrm B}}
\newcommand{\dsone}{d_{S^1}}
\newcommand{\gammaloc}{\gamma_{\mathrm{loc}}}
\newcommand{\gammaref}{\gamma_{\mathrm{ref}}}
\newcommand{\gammadrift}{\gamma_{\mathrm{drift}}}
\DeclareMathOperator{\wrap}{wrap}

\title{Quantum Shock and Train-Time Constrained Quantization:\\
A Geometric Study of Finite-Resolution Variational Quantum Circuits}

\author{Vinayak Kumar Singh\IEEEauthorrefmark{1}, Jyotsna Rani\IEEEauthorrefmark{2} and Dr. Ganesh Prasad\IEEEauthorrefmark{1} \\ 
\IEEEauthorrefmark{1}Department of Electronics and Communication Engineering,\\
National Institute of Technology Silchar, Assam, India\\
\IEEEauthorrefmark{2}Department of Electronics and Communication Engineering,\\
Indian Institute of Technology Guwahati, Assam, India}

\begin{document}
\maketitle

\begin{abstract}
Finite-resolution interfaces map Variational-Circuit angle to discrete codebook. Since Pauli rotations are periodic, so Euclidean rounding may produce wrong codeword near a chart boundary; error due to coordinates alone cannot measure how much the final state was changed. Herein we define a metric, Quantum Shock, for Pure State it is defined as the average of the square of the Fubini-Study distance (over all inputs) between the pre-deployment and deployed states. This value is limited by constant multipliers of the mean of infidelity and provides upper limits on displacement when observing bounded quantities and margin based predictive stability. Extensions to the analytical Bures-angle for mixed-states were derived, along with a finite shot stability bound. The proposed Train-Time Constrained Quantization allows a delay in applying a penalty for selecting a codeword aligned with periodicity until after task only optimizations. Three small binary tasks were evaluated on a four qubit, two layer classifier using ten matched seeds, K $\epsilon$ {4, 8, 16} and exact expectation values from simulation under noiseless conditions and state vector simulations. On our primary evaluation, T-CQ exhibited smaller average local quantum shock compared to circular post-training quantization for each task and code-book size tested. The re-execution on the x86-64 platform using K=8 preserved the direction of the mean-shock comparison but had varying magnitudes.

\end{abstract}

\begin{IEEEkeywords}
Bures angle, finite-resolution deployment, Fubini--Study distance, parameter quantization, quantum machine learning, variational quantum circuits.
\end{IEEEkeywords}

\section{Introduction}
Variational Quantum Algorithms are an integration of a parameterized quantum circuit and a classical optimization loop. They provide a wide platform for quantum simulation, chemistry, optimization and machine learning \cite{preskill2018,mcclean2016,cerezo2021, bharti2022}. Within the context of quantum machine learning, trainable rotation gates are used in a variety of quantum machine learning architectures including feature-space models, variational classifiers and circuit-learning architectures \cite{biamonte2017, schuld2019, havlicek2019, mitarai2018}. An advantage of the continuous variational formulation is that there exists a smooth parameter space over which standard optimizers can search. However, a disadvantage of this formulation is that it relies heavily upon the specific architecture of the quantum circuit, the way in which data is encoded into the parameters of the quantum circuit and the initialization of those parameters. Additionally, the number of degrees-of-freedom of the state manifold accessible by the variational algorithm plays a major role in determining its performance. There exist many variations among ansatz families regarding expressibility and entangling capabilities. Also, some highly random or redundant parameterizations will lead to either barren plateaus or directions where no change occurs in the state that is being prepared \cite{mcclean2018, haug2021, larocca2023, schuld2021encoding, fontana2022}.

Most variational models employ real-valued variables to represent the rotation angles of the circuits they are designed to prepare. It is possible for these models to be represented differently than this after deployment. For example, a compiler, controller, waveform table or memory interface could limit a rotation coordinate to a finite set of values that are allowed. This limits the problem to not just be related to hardware mechanisms such as digital control precision but also other forms such as constrained compilation, reducing control-memory traffic, a selected pulse library and/or a lower complexity deployment format. Some current processors have exposed fractional arbitrary-angle operations for example IBM document fractional $R_X(\theta)$ and bounded-angle $R_{ZZ}(\theta)$ operation support on Heron systems \cite{ibm_fractional}. Therefore, the finite codebooks we study here are determined deployment alphabets used to examine deterministic finite-resolution behavior.

There exist several previous works addressing some of the aspects of this problem. Methods incorporating compression awareness include parameter quantization incorporated into circuit compilation \cite{hu2022} while probabilistic angle interpolation uses a finite set of available rotations along with additional sampling and post-processing to estimate observables for a target angle \cite{koczor2024}. Studies involving low-resolution quantum neural networks model finite converter precision during both training and inference and use stochastic quantization to circumvent deadlocks due to gradients \cite{bhattacharjee2026}. Precision-aware analysis has also analyzed effects of analog parameter errors and digitization in quantum approximate optimization algorithms \cite{quiroz2025}. These analyses show that finite parameter precision impacts both optimization and execution. The purpose of this work is to focus on the deterministic endpoint where once a trained circuit is produced, it must map to one codeword per rotation coordinate and be executed as one fixed deployed circuit.

At this endpoint, three connected questions emerge. What codeword(s) would be intrinsically closest when the parameter is periodic? How should the physical effects of resulting projections be measured when coordinate changes that are equally sized can cause differing size state changes? Can the optimization trajectory be generated to allow the hard projection at its end to be minimized? To answer these questions requires both parameter-space and state-space geometries.

For a Pauli operator $p$,
The rotations are defined by
\begin{equation}
R_P(\theta)=e^{-i\theta p/2},\\
R_p(\theta+2\pi)=-r_p(\theta)
\label{eq:pauli_rotation}.
\end{equation}
The global phase in Eq. \eqref{eq:pauli_rotation} leads to an independent Pauli rotation coordinate that is $2\pi$ periodic in projective Hilbert space. Circuits composed of such rotations with $n$ individually controlled parameters can be naturally described on the product domain $\mathbb{T}^N=(S^1)^N$. An unwrapped interval remains a valid coordinate chart but it places physically adjacent angles near $-\pi$ and $\pi$ at opposite numerical endpoints. Ordinary scalar roundings performed within those charts will often choose a point on the "long" arc of the unit-circle. Therefore, we need to use circular nearest codewords to recover the lost information due to that boundary mismatch.   
 
A single topology correction will not capture all effects of damage incurred during the process of deploying a quantum circuit. In addition to being able to represent complex spatial relationships within their parameters, encoding circuits can be highly directional; they can have redundant coordinate representations; they can include many symmetry elements; and their responses to changes in state can be highly anisotropic \cite{haug2021,larocca2023,schuld2021encoding,fontana2022}. Therefore, two different parameter variations that are identical in terms of their circular norms could potentially result in vastly different output states or predictions when encoded into the circuit. As such, we consider it beneficial to evaluate how well a circuit performed at each point in time after the application of the circuit mapping in projective Hilbert space. Here, for our pure-state simulation results, we define Quantum Shock as the expected square of the Fubini-Study distance between a method’s pre-deployment state and its final deployed state \cite{provost1980,anandan1990,braunstein1994}. Extensions to density operator-based methods are achieved by using the Bures angle, which is derived from the square root of the Uhlmann fidelity. Pure-states and Bures angles will yield equivalent results \cite{fuchs1999,nielsen2010}. Additionally, this framework tracks both how far away a continuously chosen reference is from the final deployed state, and the total amount of displacement that occurred throughout the course of training over time. Distinguishing these two types of displacement allows us to determine if the robustness against the final hard-mapping was due solely to the topology of the original continuous checkpoint, or if it also reflected the degree to which the model was trained towards the discrete codebook.

Our experimental methodology follows \emph{Unwrap Euclidean Projection} → \emph{Circular PTQ} → \emph{T-CQ}. We compare the performance of Circular PTQ to T-CQ. The former maps a validation-selected Continuously sampled checkpoint onto a phase zero circular codebook without retraining. T-CQ is similar to Circular PTQ in that it utilizes the same hard codebook and Euclidean projection rule. However, while Circular PTQ utilized the full optimization trajectory to achieve the desired deployment state, T-CQ restricted itself to optimize for only a portion of the training trajectory until activation of a normalized periodic penalty, which occurs once a certain percentage of training has taken place. When activated, the penalty minima occur at the points in time where the deployment codewords were identified. After validation data selection for both the raw checkpoint and one of eleven predetermined training schedule configurations, a deterministic projection of the circuit is made onto the selected configuration and then tested on the held out test set. The primary experiment used ten matches of seeds, three binary classification problems, and K = \{4, 8, 16\} angular configurations per dimension. An additional fully executed version of this study was conducted using a different processor and software build to ensure that the main orderings of displacement persisted across different numerical environments.

This work includes four contributions:
\begin{enumerate} 
\item We formulate deterministic nearest-codeword projection for periodic Pauli-rotation coordinates and distinguish local deployment shock, training-path displacement, and displacement from an independently selected continuous reference.

\item The relationship between Squared Fubini-Study Distance and Mean Infidelity is established and derive observable, margin-stability, mixed-state, finite-shot, and coordinate-to-state bounds under stated assumptions.

\item We introduce T-CQ (a Delayed Codebook-Aligned Periodic-Hardening Procedure) which has a specified Schedule, Validation Selection Rule and Deterministic Deployment Map.

\item We report Accuracy and State Displacement for the declared Baseline(s), Three Small Binary Tasks, Ten Matched Seeds and K $\epsilon$ {4,8,16} separately from the Primary Run and including a Separate x86-64 Execution.
\end{enumerate}

The remainder of the sections are arranged as follows. In Section \ref{related_work} we review the literature on related work, and Section \ref{finite-resolution-deployment} discusses finite-resolution quantum deployment. Section \ref{geometric-state-space} introduces quantum state-geometry. In Section \ref{geometry} we develop the periodic placement formulations and the Quantum Shock measures. In Section \ref{robustness} we establish the robustness results. In Section \ref{tcq} we introduce T-CQ. In Section \ref{setup} we describe how we will simulate the system. In Sections \ref{results} and \ref{discussion} we present and discuss our results. Finally, we conclude with Section \ref{conclusion}.

\section{Related Work}\label{related_work}

\subsection{Variational Circuits and Quantum Learning}
Using a parametrizable circuit as a trainable model, where the parameters are optimized by a classical optimizer, has been explored using Variational Quantum Algorithms \cite{preskill2018,mcclean2016,cerezo2021,bharti2022}. In particular, applications have emerged such as eigensolvers, quantum simulation, combinatorial optimization, feature mapping and supervised learning. Models in quantum-learning are able to encode data in terms of rotation angles or feature maps, while using trainable layer of the circuit to produce a scalar observable (or kernel) \cite{biamonte2017,schuld2019,havlicek2019,mitarai2018}. The structure of the circuit defines what states and functions are possible to reach; on the other hand, the optimization landscape describes how easy it is to reach those states.

Expressiveness, ability to create entanglement and trainability are not identical. Random or sufficiently expressible circuits may develop a "barren plateau" \cite{mcclean2018} where useful gradients will become exponentially small. The connection between trainability and expressivity can be described in terms of the geometric analysis of quantum systems as effective dimensions and ranks of the parameter space \cite{haug2021}. While overparametrization can provide better access to relevant state-space direction; the parameters beyond the effective rank do not add anything \cite{larocca2023}. Gates encoding data define the Fourier spectrum available to a quantum learning model \cite{schuld2021encoding}; whereas symmetries within the circuit allow identification of distant points in the parameter space with respect to both physical states \cite{fontana2022} and objective function values. This motivates viewing a variational circuit as a mapping from parameters to physical states rather than as each coordinate displacements being an equally valid change.

This work takes place at the deployment limit of this continuous variational framework. The traditional four qubit variational classifier architecture was used in order to isolate the effects of finite angular deployment from variations in circuit capacity.

\subsection{Quantization and Periodic Regularization}
Scalar quantization is a well known technique from classical signal processing to map real values to a discrete set of values according to some distortion metric \cite{gray1998}. When one applies post-training quantization, the continuous model is trained prior to projecting it into the discrete set. In contrast, when one applies quantization-aware training, the continuous model is exposed to the discrete representation during training \cite{bengio2013}. One common approach to achieve this exposure is by utilizing straight-through estimators. During the forward pass either a hard or stochastic quantizer is applied; whereas during the backward pass a surrogate derivative is substituted for the original quantizer.

Periodic regularization is an example of a differentiable alternative. Naumov et al. investigated sine, cosine and hat-shaped penalties that have their minima attracting classical neural network weights towards discrete values \cite{naumov2018periodic}. SinReQ provided a general sinusoidal regularizer for low bit width training and also provided flexibility for having different quantization levels per layer \cite{elthakeb2019sinreq}. The authors demonstrate that periodic potentials can prepare a continuous optimizer for a discrete end point without requiring the application of a hard projection throughout all of training.

T-CQ\ specializes periodic regularization to circular Pauli-rotations. The number and phase of the wells were determined so that the minimum of the potential coincides with the angular codebook to be employed during deployment. A normalized schedule delays the penalty until after an initial task-optimization phase and reaches its selected maximum at the final epoch. The method is compared to a matched circular post-training projection, allowing separation of contributions due to train-time preparation versus topology of the final quantizer.

\section{Finite-Resolution Quantum Deployment}\label{finite-resolution-deployment} Methods for the specific quantization of the deployment object vary. CompVQC unifies prunings, parameter quantizations, and compilation objectives to lower variational circuit complexity \cite{hu2022}. QuantumNAT integrates noise addition and normalization with post-measurement output quantization \cite{wang2022quantumnat}; however, it is the measured features (not the trainable rotation coordinates used here) that serve as the quantized objects.

Probabilistic Angle Interpolation (PAI) substitutes discrete rotation sets in place of the continuous angle being approximated \cite{koczor2024}. PAI samples three discrete rotation settings and combines their measurements to estimate the expected value corresponding to a target continuous angle. As such, this approach trades off deterministic unitary approximation with sampling and post-processing. T-CQ\ operates at a different point; each coordinate is assigned a single codeword and the resultant "hard" circuit is the deployed model.

Bhattacharjee et. al. represent the most directly relevant studies of control resolution \cite{bhattacharjee2026}. They investigate both training and inference under finite D/A converter resolution and demonstrate how using temperature-controlled stochastic quantization can eliminate deadlocks due to gradient vanishing. Both classifiers share an architecturally material overlap: four qubits, $R_y$ feature encoding, two layers of trainable $R_y$ and $R_z$ rotations, a circular CNOT ring, sixteen trainable angles, and a first-qubit Pauli-Z score. Therefore, we do not claim that the current work contributes anything architecturally to this shared architecture. Our questions and protocols were different from those in Bhattacharjee et. al.: they discretize inputs and parameters to a closed interval of converter steps and evaluate stochastic discrete updates and accuracy; our approach uses distinct points on a circular codebook, separates circular post-training projection from delayed deterministic hardening, and investigates projective-state displacement. We make no claims regarding relative superiority to their stochastic method without reproducing results comparably.

Variational optimization outside of quantum neural networks also includes finite precision. Quiroz et. al. establish performance limits for finite precision errors in the QAOA and illustrate that analogous error reduction can be achieved through increased circuit depth by utilizing a coarser or synthesized representation of the parameters \cite{quiroz2025}. This demonstrates a more general trade-off: a more precise or synthesized representation may decrease error in parameters but increase some other deployment resource.

\subsection{Gate Synthesis, Control and Execution Errors/Noise}

Fault-tolerant synthesis represents an alternative path to discrete implementations based on continuous rotations. Methods such as Clifford + T provide an example of how to approximate any logical rotation using sequences of gates of varying lengths depending upon desired accuracy \cite{ross2016}. However, instead of mapping each trainable coordinate to a sequence of gates in the case of the architecture examined in this paper, we map each trainable coordinate to one of a predetermined set of angular settings.

Thus, there exists a difference in the type of deployment objects utilized in these two approaches: a sequence of discrete gates in synthesis and a single finite-resolution setting in the case of codebook projection.

Finally, at the physical-control level, the ultimate determinate of the angle of a circuit is a pulse, waveform, duration, amplitude, or phase. Optimal control theories have emphasized that variational parameters represent limited control resources, and that both circuit-level and pulse-level optimizations are related \cite{magann2021}. Nonidealities of classical controllers can also impose quantitative requirements on the fidelity of qubit operations and readouts \cite{vandijk2019}. Thus, even if the codebook itself is established at either the model or compiler level, finite-resolution deployments represent a hardware-software interface problem.

Circular deterministic parameter projections are also different than quantum error mitigation techniques. For instance, zero-noise extrapolation and quasi-probability methods address biased expectations due to noise in circuit execution \cite{temme2017}, while model-free readout mitigations target biases in measured expectation values due to noise in measurement \cite{vandenberg2022}. These types of mitigations occur after the circuit has been defined. Circular projections and T-CQ\ occur before the definition of the circuit; thus, the mechanisms are complementary: one reduces error in the representation of the deployed model; the other reduces error added during execution or measurement.

\section {Quantum-State Geometry}\label{geometric-state-space}
Pure quantum states represent rays (rays) and not vectors with globally meaningful phases, as these do not have physical meanings. The Fubini-Study metric defines the intrinsically defined geodesic distances between rays on a pure projective state space \cite{provost1980, anandan1990}. The squared Uhlmann fidelity and the associated Bures angle generalize ray-based comparisons to density operators \cite{braunstein1994, fuchs1999, nielsen2010}, while trace distance measures statistical distinguishability and variation in bounded observables. The real part of the quantum geometric tensor represents the pullback of the Fubini-Study geometry onto the circuit-parameter manifold and is related to quantum natural gradient approaches \cite{stokes2020} and can be used to identify insensitive/overlapping parameters \cite{haug2021, larocca2023}.

A key element separating coordinate geometry from state geometry is critical when considering finite resolution deployments. While circular distance determines which codeword is closest to a point on an independent Pauli rotation coordinate, it does not quantify how much the output state has changed. Quantum shock introduces the analogous state space measure. Its squared-Fubini-Study form corresponds to mean-infideliity within universal constants. It satisfies a useful geodesic-triangle inequality. Finally, it maps paths in parameter space onto paths in observable displacement. The Bures-angle generalization maintains the equivalent-infidelity relationship and observable control over mixed-output states, without claiming that the simulations provided contain a noise channel.

In summary, this work unifies three existing concepts into a deployment specific context: circular nearest-codeword projection for periodic coordinates, a projective-state displacement function, and delay-codebook aligned hardening. Circular PTQ provides the topology-only control, while T-CQ\ addresses whether training prepares a raw checkpoint that is locally stable under the same deterministic projection.

\section{Geometric Formulation}\label{geometry}

\subsection{Periodic Parameter Domain}
Let $\mathcal H=(\mathbb C^2)^{\otimes n}$ and
\begin{equation}
\ket{\psi(x;\bm\theta)}=U(x,\bm\theta)\ket{0}^{\otimes n},
\qquad \bm\theta\in\mathbb T^N,
\label{eq:model_state}
\end{equation}
where $x$ follows an evaluation distribution and $\mathbb T^N=(S^1)^N$. This specifies the parameter domain, not the topology of the circuit image: symmetries can identify multiple parameter vectors with the same projective state.
The product-circle model applies here to independently parameterized, uncontrolled Pauli rotations. Controlled or multi-qubit parameterized gates can possess different periods or additional quotient identifications and must be analyzed using their own gate family.

\begin{figure}[t]
\centering
\includegraphics[width=0.94\columnwidth]{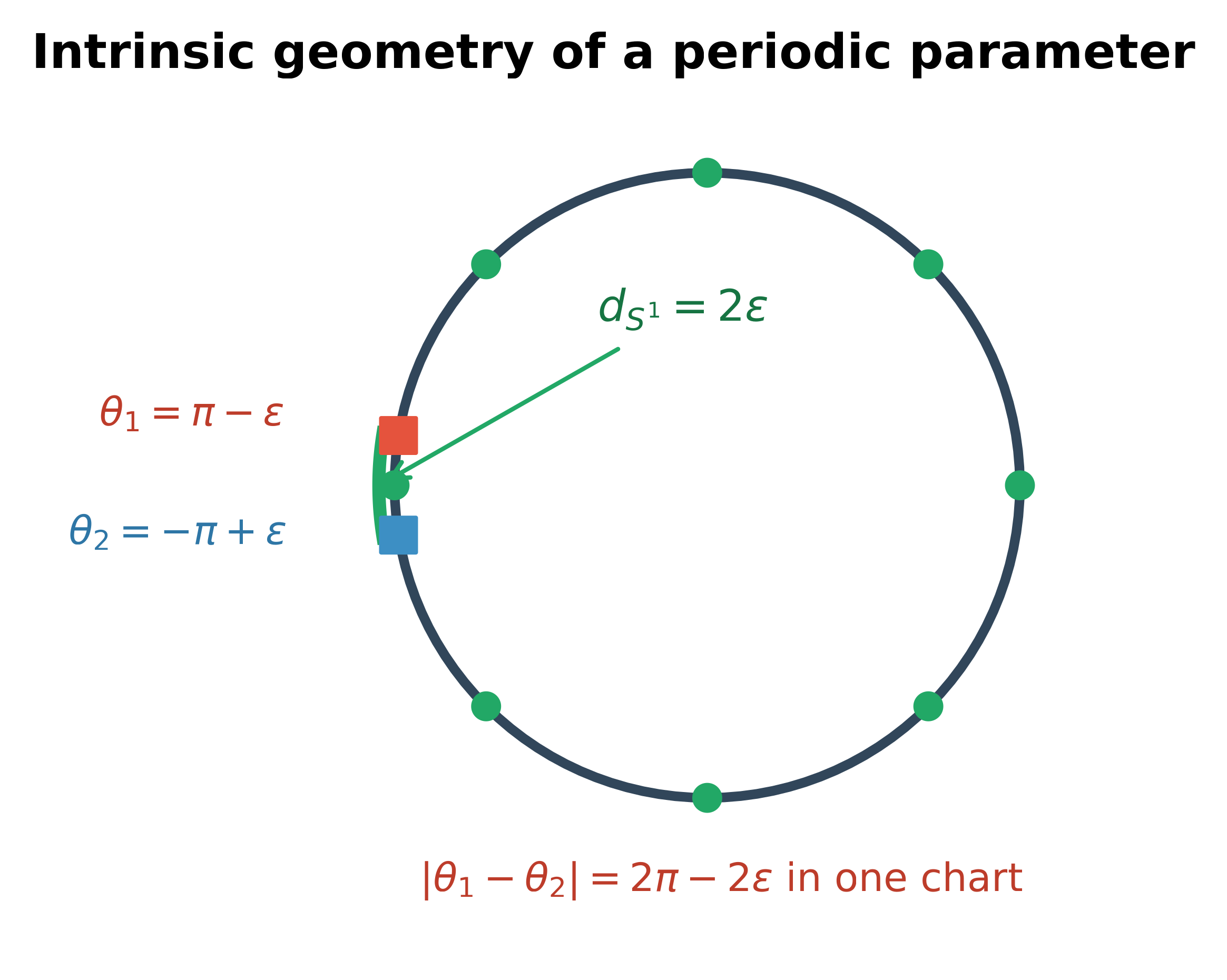}
\caption{A periodic coordinate near the chart boundary. Green markers show the common phase-zero $K=8$ deployment codebook used by Circular-PTQ and T-CQ. Circular distance selects along the short arc, whereas an unwrapped scalar difference crosses the chart and can select a physically more distant codeword.}
\label{fig:geometry}
\end{figure}

For representatives in $[-\pi,\pi)$, define
\begin{equation}
\dsone(\theta,\phi)
=\left|\wrap_{[-\pi,\pi)}(\theta-\phi)\right|.
\label{eq:circular_distance}
\end{equation}
Thus $\theta_1=\pi-\epsilon$ and $\theta_2=-\pi+\epsilon$ have circular distance $2\epsilon$, although their difference in that chart is $2\pi-2\epsilon$; Fig.~\ref{fig:geometry} illustrates the mismatch.

For an ordered finite codebook
$\mathcal C_i=(c_{i,0},\ldots,c_{i,K_i-1})\subset S^1$, define the exact nearest-codeword map by
\begin{equation}
\begin{aligned}
d_i^\star(\theta_i)
&=\min_{0\leq j<K_i}\dsone(\theta_i,c_{i,j}),\\
j_i^\star(\theta_i)
&=\min\!\left\{j:\dsone(\theta_i,c_{i,j})=d_i^\star(\theta_i)\right\},\\
Q_i(\theta_i)&=c_{i,j_i^\star(\theta_i)}.
\end{aligned}
\label{eq:quantizer}
\end{equation}
Thus an exact Voronoi tie is resolved by the smallest stored codeword index. The exact analysis below uses $Q_i$. Floating-point offline projection uses the explicitly distinct $\varepsilon_{\rm tie}$-nearest map
\begin{equation}
\begin{aligned}
\widehat Q_i(\theta_i)&=c_{i,\widehat j_i(\theta_i)},\\
\widehat j_i(\theta_i)&=\min\!\left\{j:\;
\dsone(\theta_i,c_{i,j})\leq
d_i^\star(\theta_i)+\varepsilon_{\rm tie}\right\}.
\end{aligned}
\label{eq:numerical_quantizer}
\end{equation}
with $\varepsilon_{\rm tie}=10^{-12}$ rad. This numerical convention is not asserted to be exact nearest-neighbor projection.

For an integer $K\geq2$, the experiments use the ordered phase-zero codebook
\begin{equation}
\begin{aligned}
\mathcal C_K&=(c_0,\ldots,c_{K-1}),\\
c_j&=\wrap_{[-\pi,\pi)}\!\left(\frac{2\pi j}{K}\right),
\quad j=0,\ldots,K-1.
\end{aligned}
\label{eq:codebook}
\end{equation}
The reported values use $K\in\{4,8,16\}$; Fig.~\ref{fig:geometry} shows the $K=8$ case with spacing $\Delta_\theta=\pi/4$. Every uniform $K$-level circular codebook is a rotation of Eq. \eqref{eq:codebook}. Phase zero fixes one common, nonoptimized reference alphabet for all methods and prevents codebook offset from becoming an additional selected hyperparameter. It is not claimed to be universally optimal; a calibrated interface could instead supply fixed offsets $\phi_i$, which are supported by the regularizer in Eq. \eqref{eq:regularizer}.

\subsection{Assumptions and Scope}
The product-circle parameterizations in \cref{eq:circular_distance,eq:quantizer,eq:numerical_quantizer,eq:codebook} along with the nearest codeword rule are based upon independently parameterized, uncontrolled Pauli rotation gates with a period of $2\pi$ within the projective Hilbert space. Additionally, the bound provided in Proposition \ref{prop:coordinate_bound} applies when each parameter appears exactly once in the quantum circuit. No claims about bounds are made regarding multiple or shared parameters, control of the rotations, or other gate families that have different periods or generators. Each would need their own parameter domain and a new bound. The results regarding observable outcomes in Section \ref{robustness} depend on the assumption of an output observable being Hermitian-bounded and the output map being Lipschitz; the margin result relies upon a specified distribution of scores to margins. The result regarding finite shots in this section is based on independence among projective measurement outcomes. The Bures angle results provide an analytical extension to the distance-based analyses of mixed states. However all experiments presented here rely on noise-free simulations using pure states and exact expectations.

\subsection{State-Space Quantities}
For normalized pure states, use
\begin{equation}
\dfs(\psi,\phi)=\arccos|\langle\psi|\phi\rangle|,
\qquad 0\leq\dfs\leq\frac{\pi}{2}.
\label{eq:fs_distance}
\end{equation}
For any two checkpoints $\bm\theta$ and $\bm\phi$, define
\begin{equation}
\gamma(\bm\theta,\bm\phi)
=\mathbb E_x\!\left[\dfs^2\!\left(
\psi(x;\bm\theta),\psi(x;\bm\phi)\right)\right].
\label{eq:general_gamma}
\end{equation}

Let $F_x=|\langle\psi(x;\bm\theta)|\psi(x;\bm\phi)\rangle|^2$ and define the mean infidelity $\bar\iota=\mathbb E_x[1-F_x]$.
\begin{proposition}[Quantum Shock and infidelity]
\label{prop:infidelity}
For pure states,
\begin{equation}
\bar\iota=\mathbb E_x[\sin^2 d_x],\qquad
\boxed{\bar\iota\leq\gamma(\bm\theta,\bm\phi)
\leq\frac{\pi^2}{4}\bar\iota},
\label{eq:infidelity_equivalence}
\end{equation}
where $d_x=\dfs(\psi(x;\bm\theta),\psi(x;\bm\phi))$.
\end{proposition}
\begin{proof}
By definition, $d_x=\arccos\sqrt{F_x}$, hence $1-F_x=\sin^2d_x$. For $d_x\in[0,\pi/2]$, $(2/\pi)d_x\leq\sin d_x\leq d_x$. Squaring and averaging gives Eq. \eqref{eq:infidelity_equivalence}.
\end{proof}

\paragraph*{Mixed-state extension.}
For density operators $\rho$ and $\sigma$, use the squared Uhlmann fidelity and Bures angle
\begin{align*}
F_{\rm U}(\rho,\sigma)
&=\left(\operatorname{Tr}\sqrt{\sqrt{\rho}\,\sigma\sqrt{\rho}}\right)^2,\\
\abures(\rho,\sigma)
&=\arccos\sqrt{F_{\rm U}(\rho,\sigma)},
\qquad 0\leq\abures\leq\frac{\pi}{2}.
\end{align*}
For state families $\{\rho_x\}$ and $\{\sigma_x\}$, define
\begin{align*}
\gammab(\rho,\sigma)
&=\mathbb E_x[\abures^2(\rho_x,\sigma_x)],\\
\bar\iota_{\rm B}
&=\mathbb E_x[1-F_{\rm U}(\rho_x,\sigma_x)].
\end{align*}
Because $1-F_{\rm U}=\sin^2\abures$, the proof of Proposition~\ref{prop:infidelity} applies without change and gives
\[
\bar\iota_{\rm B}\leq\gammab(\rho,\sigma)
\leq\frac{\pi^2}{4}\bar\iota_{\rm B}.
\]
For pure states, $F_{\rm U}=|\langle\psi|\phi\rangle|^2$, $\abures=\dfs$, and $\gammab=\gamma$. Thus $\gammab$ is the mixed-state extension of the same Quantum Shock convention, rather than a different empirical metric. Since the Bures angle is a metric, replacing $\dfs$ by $\abures$ also preserves the triangle-and-Minkowski bound in Eq. \eqref{eq:gamma_decomposition} for mixed-state families.
Thus Quantum Shock is equivalent to mean infidelity up to a universal factor. We use squared geodesic distance because $d_{\rm FS}$ is intrinsic, obeys the triangle inequality, and connects directly to parameter-manifold path lengths; this choice does not imply that fidelity is inadequate.

Let $\bm\theta_M^{\rm raw}$ and $\bm\theta_M^{\rm dep}$ denote method $M$ immediately before and after deployment, and let $\bm\theta_{\rm ref}$ be the independently validation-selected continuous model. We report
\begin{align}
\gammaloc(M)
&=\gamma(\bm\theta_M^{\rm raw},\bm\theta_M^{\rm dep}),
\label{eq:gamma_local}\\
\gammaref(M)
&=\gamma(\bm\theta_{\rm ref},\bm\theta_M^{\rm dep}).
\label{eq:gamma_reference}
\end{align}
We call $\gammaloc$ the method's \emph{Quantum Shock}. Define also the training-path displacement
\begin{equation}
\gammadrift(M)=\gamma(\bm\theta_{\rm ref},\bm\theta_M^{\rm raw}).
\label{eq:gamma_drift}
\end{equation}


The "Quantum Shock" for a specific method locally is the average, taken with respect to the evaluation input(s), of the squared Fubini-Study angle between the state created at deterministic deployment and the state produced by codebook projection. In terms of units, it is measured as the square of a radian: if the codebook projection has no effect on the state, then the shock will be zero. A larger shock indicates greater average displacement of the state from its initial position.

Eq. \ref{eq:infidelity_equivalence} establishes a relationship between Quantum Shock and expected infidelity that includes all universal constants. Quantum Shock is a measure of state displacement and not of task utility. Hence, a lower shock does not guarantee better classification performance. Better classification can depend upon the size of the margins associated with each classification score, as well as on whether or not the state displacement is directed towards or away from the decision boundary.

The pointwise triangle inequality for $\dfs$, followed by Minkowski's inequality in $L^2$, yields the upper bound
\begin{equation}
\sqrt{\gammaref(M)}
\leq \sqrt{\gammadrift(M)}+\sqrt{\gammaloc(M)}.
\label{eq:gamma_decomposition}
\end{equation}
This is not an additive decomposition of $\gammaref$: slack can enter both the pointwise triangle inequality and Minkowski's inequality. Consequently, the magnitudes of $\gammadrift$ and $\gammaloc$ do not identify fractional or causal contributions to $\gammaref$. For a post-training quantizer, $\bm\theta_M^{\rm raw}=\bm\theta_{\rm ref}$, so $\gammadrift=0$ and $\gammaloc=\gammaref$. For train-time methods, $\gammaloc$ measures robustness to the method's own final deployment map, $\gammadrift$ measures optimization-path displacement before deployment, and $\gammaref$ measures end-to-end displacement.

A hard-forward surrogate method requires one additional distinction. It stores latent parameters $\bm\vartheta$ while training uses a hard-quantized forward circuit. For a common parameter-map diagnostic we define
\begin{equation}
\gamma_{\rm map}=\gamma(\bm\vartheta,\widehat Q(\bm\vartheta)).
\label{eq:gamma_map}
\end{equation}
For Continuous, post-training projection, and T-CQ, the stored parameters are also the raw forward parameters, so $\gamma_{\rm map}=\gammaloc$. For the archived STE-QAT execution, tensor \texttt{argmin} implements the exact map during gradient-bearing forward passes, whereas the offline validation and export path uses $\widehat Q$. Therefore $\gamma_{\rm map}$ is a latent-to-export diagnostic, not a measured deployment switch, and no zero-switching-shock claim is made for numerical near ties. The reported $\gammaref$ compares the independent continuous reference with the exported hard circuit.

\section{Robustness Theory}\label{robustness}

Suppose the model output is a bounded-observable expectation followed by a classical map,
\begin{equation}
z_{\bm\theta}(x)=\operatorname{Tr}[O\rho_x(\bm\theta)],
\quad s_{\bm\theta}(x)=g(z_{\bm\theta}(x)),
\label{eq:score}
\end{equation}
where $O$ is Hermitian, $g$ is $L$-Lipschitz, and
$\rho_x(\bm\theta)$ is a density operator. For two checkpoints, write
$A_x=\abures(\rho_x(\bm\theta),\rho_x(\bm\phi))$ and
$\gammab(\bm\theta,\bm\phi)=\mathbb E_x[A_x^2]$; for pure states these reduce to $d_x$ and $\gamma(\bm\theta,\bm\phi)$, respectively. Write $\operatorname{osc}(O)=\lambda_{\max}(O)-\lambda_{\min}(O)$ for the spectral width of $O$. Theorem~\ref{thm:observable} and Corollary~\ref{cor:margin} concern exact expectation values; Corollary~\ref{cor:shots} adds direct finite-shot measurement.

\begin{theorem}[Quantum-Shock observable bounds]
\label{thm:observable}
For any checkpoints $\bm\theta,\bm\phi$ and every input $x$,
\begin{equation}
\begin{aligned}
|s_{\bm\theta}(x)-s_{\bm\phi}(x)|
&\leq L\operatorname{osc}(O)D_x\\
&\leq L\operatorname{osc}(O)\sin A_x
\leq L\operatorname{osc}(O)A_x,
\end{aligned}
\label{eq:pointwise_bound}
\end{equation}
where $D_x=\frac12\|\rho_x(\bm\theta)-\rho_x(\bm\phi)\|_1$. Consequently,
\begin{align}
\mathbb E_x|s_{\bm\theta}-s_{\bm\phi}|
&\leq L\operatorname{osc}(O)\sqrt{\gammab(\bm\theta,\bm\phi)},
\label{eq:mean_bound}\\
\mathbb E_x|s_{\bm\theta}-s_{\bm\phi}|^2
&\leq L^2\operatorname{osc}^2(O)\gammab(\bm\theta,\bm\phi).
\label{eq:mse_bound}
\end{align}
\end{theorem}

\begin{proof}
Set $\rho=\rho_x(\bm\theta)$ and $\sigma=\rho_x(\bm\phi)$.
The Fuchs--van de Graaf inequality gives
$D_x\leq\sqrt{1-F_{\rm U}(\rho_x(\bm\theta),\rho_x(\bm\phi))}
=\sin A_x$ \cite{fuchs1999,nielsen2010}.
Let $c=[\lambda_{\max}(O)+\lambda_{\min}(O)]/2$. Since $\operatorname{Tr}(\rho-\sigma)=0$, trace-norm/operator-norm duality gives
$|\operatorname{Tr}[O(\rho-\sigma)]|
=|\operatorname{Tr}[(O-cI)(\rho-\sigma)]|
\leq\frac12\operatorname{osc}(O)\|\rho-\sigma\|_1
=\operatorname{osc}(O)D_x$.
Lipschitz continuity, $D_x\leq\sin A_x$, and $\sin A_x\leq A_x$ establish
Eq. \eqref{eq:pointwise_bound}. Cauchy--Schwarz gives
$\mathbb E A_x\leq\sqrt{\mathbb E A_x^2}$, proving
Eq. \eqref{eq:mean_bound}; squaring the pointwise inequality and averaging proves
Eq. \eqref{eq:mse_bound}.
\end{proof}

For the measured Pauli observable $Z_0$, $\operatorname{osc}(O)=2$ and the experiment uses $g(z)=z$, so $L=1$.

\begin{corollary}[Margin-distribution stability]
\label{cor:margin}
Let $b\in\mathbb R$ be the classification threshold,
$h_{\bm\theta}(x)=\mathbf 1\{s_{\bm\theta}(x)\geq b\}$, and
$m_{\bm\theta}(x)=|s_{\bm\theta}(x)-b|$. For every $a>0$,
\begin{equation}
\begin{aligned}
\Pr[h_{\bm\theta}\neq h_{\bm\phi}]
\leq\min\!\Bigg\{1,\;&\Pr[m_{\bm\theta}<a]\\
&+\frac{L^2\operatorname{osc}^2(O)}{a^2}
\gammab(\bm\theta,\bm\phi)\Bigg\}.
\end{aligned}
\label{eq:margin_distribution}
\end{equation}
If $m_{\bm\theta}(x)\geq m>0$ almost surely, the first term vanishes at $a=m$.
\end{corollary}

\begin{proof}
Outside the event $m_{\bm\theta}<a$, a label change requires
$|s_{\bm\theta}-s_{\bm\phi}|\geq a$. The union bound, Markov's inequality applied to the squared score displacement, Eq. \eqref{eq:mse_bound}, and the trivial probability upper bound one give the result.
\end{proof}

\begin{corollary}[Finite-shot deployment stability]
\label{cor:shots}
Let $\{\rho_x\}$ be a reference state family with exact score
$s_\rho(x)=g(\operatorname{Tr}[O\rho_x])$, and let $\{\sigma_x\}$ be a deployed state family. Suppose $\widehat z_\sigma(x)$ is the sample mean of $M\geq1$ independent projective measurements of $O$ on $\sigma_x$, and set $\widehat s_\sigma(x)=g(\widehat z_\sigma(x))$. With expectation over inputs and measurement outcomes,
\[
\left(\mathbb E_{x,{\rm shot}}
|s_\rho-\widehat s_\sigma|^2\right)^{1/2}
\leq L\operatorname{osc}(O)
\left(\sqrt{\gammab(\rho,\sigma)}+\frac{1}{2\sqrt M}\right).
\]
If $h_\rho=\mathbf1\{s_\rho\geq b\}$,
$\widehat h_\sigma=\mathbf1\{\widehat s_\sigma\geq b\}$, and
$m_\rho=|s_\rho-b|$, then for every $a>0$,
\[
\begin{aligned}
\Pr[h_\rho\neq\widehat h_\sigma]
\leq\min\!\Bigg\{1,\;&\Pr[m_\rho<a]\\
&+\frac{L^2\operatorname{osc}^2(O)}{a^2}\\[-0.2ex]
&\quad\times
\left(\sqrt{\gammab(\rho,\sigma)}
+\frac{1}{2\sqrt M}\right)^2\Bigg\}.
\end{aligned}
\]
\end{corollary}

\begin{proof}
A projective measurement of $O$ has outcomes in
$[\lambda_{\min}(O),\lambda_{\max}(O)]$. Hence its variance is at most $\operatorname{osc}^2(O)/4$, and the conditional mean-square error of the $M$-shot sample mean is at most $\operatorname{osc}^2(O)/(4M)$. Lipschitz continuity and Minkowski's inequality in $L^2$ give
\begin{align*}
\|s_\rho-\widehat s_\sigma\|_2
&\leq L\|z_\rho-z_\sigma\|_2
+L\|z_\sigma-\widehat z_\sigma\|_2\\
&\leq L\operatorname{osc}(O)
\left(\sqrt{\gammab(\rho,\sigma)}+\frac{1}{2\sqrt M}\right).
\end{align*}
The margin statement follows from the same union-bound and squared-Markov argument as Corollary~\ref{cor:margin}.
\end{proof}

For $O=Z_0$ and $L=1$, Corollary \ref{cor:shots} the finite-shot root-mean-square bound is $2\sqrt{\gammab}+M^{-1/2}$. This is an analytical sampling guarantee; the reported tables use exact state-vector expectations rather than finite-shot estimates.

The corollary is intentionally a stability statement, not a direct accuracy guarantee: labels can remain unchanged and still be wrong, and accuracy also depends on the reference classifier.

\begin{proposition}[Coordinate-to-state bound]
\label{prop:coordinate_bound}
Assume each coordinate $\theta_i$ occurs once in a Pauli rotation $e^{-i\theta_iP_i/2}$. Let
$\delta_i=\wrap_{[-\pi,\pi)}(\phi_i-\theta_i)$. Then, for every $x$,
\begin{equation}
\dfs(\psi(x;\bm\theta),\psi(x;\bm\phi))
\leq\frac12\sum_{i=1}^{N}|\delta_i|.
\label{eq:coordinate_state_bound}
\end{equation}
For nearest projection to the uniform $K$-level codebook,
\begin{equation}
\gamma(\bm\theta,Q(\bm\theta))
\leq\min\!\left\{\frac{\pi^2}{4},
\frac{N^2\pi^2}{4K^2}\right\}.
\label{eq:uniform_global_bound}
\end{equation}
\end{proposition}

\begin{proof}
Connect $\bm\theta$ to $\bm\phi$ by changing one coordinate at a time along its shortest circular arc. For a Pauli rotation, the Fubini--Study speed is the standard deviation of the generator $P_i/2$, which is at most $\|P_i/2\|_\infty=1/2$. The length of this path is therefore at most $\frac12\sum_i|\delta_i|$, and geodesic distance cannot exceed path length. Nearest uniform projection gives $|\delta_i|\leq\pi/K$; squaring and using the universal bound $\dfs\leq\pi/2$ proves Eq. \eqref{eq:uniform_global_bound}.
\end{proof}

Equation~Eq. \eqref{eq:uniform_global_bound} is a deterministic worst-case bound for nearest projection onto the uniform codebook; it assumes neither random parameters nor random projection.

For the numerical map $\widehat Q$ in Eq. \eqref{eq:numerical_quantizer},
$\dsone(\theta_i,\widehat Q_i(\theta_i))\leq\pi/K+\varepsilon_{\rm tie}$. The same path argument therefore gives
\begin{equation}
\gamma(\bm\theta,\widehat Q(\bm\theta))
\leq
\min\!\left\{\frac{\pi^2}{4},
\frac{N^2}{4}\left(\frac{\pi}{K}+\varepsilon_{\rm tie}\right)^2\right\}.
\label{eq:numerical_uniform_bound}
\end{equation}
For the experimental circuit we have N = 16, and K $\in$ {4, 8, 16} for the bound of Eq. \ref{eq:uniform_global_bound}: $\min \left\{ \frac{\pi^2}{4}, \frac{N^2\pi^2}{4K^2} \right\}$ is approximately equal to $\pi^2/4 \approx 2.467$ for each of the three codebook sizes, and that the $\epsilon_{tie}$ correction of Eq. \ref{eq:numerical_uniform_bound} is negligible at the given level of precision. That is a valid although noninformative worst-case bound for this set of configurations. For instance, at K = 8 the average of ten local shocks per seed for T-CQ on Two Moons, Iris, and MNIST 3 vs. 6 are 0.0110, 0.0294, and 0.0104, respectively, as opposed to 0.1428, 0.1281, and 0.1153 for Circular-PTQ.
This worst-case bound can be loose because it ignores cancellations, parameter redundancies, the input distribution, and the local quantum metric. The empirical state-space quantity in Eq. \eqref{eq:general_gamma} retains those effects.

\section{Train-Time Constrained Quantization}\label{tcq}

\subsection{Codebook-Aligned Hardening}
Let $T\geq2$ and index epochs by $t\in\{0,\ldots,T-1\}$. Choose
\begin{equation*}
\begin{aligned}
0\leq\tau&<\frac{T-1}{T},\qquad
t_{\rm h}=\lfloor\tau T\rfloor\in\{0,\ldots,T-2\},\\
r&>0,\qquad \lambda_{\max}\geq0.
\end{aligned}
\end{equation*}
Then T-CQ\ uses
\begin{equation}
\mathcal L_t(\bm\theta)=\mathcal L_{\rm task}(\bm\theta)
+\lambda_t\mathcal L_{\rm geom}(\bm\theta),
\label{eq:tcq_loss}
\end{equation}
where $\lambda_t=0$ for $t<t_{\rm h}$ and, for $t_{\rm h}\leq t\leq T-1$,
\begin{equation}
\lambda_t=\lambda_{\max}
\frac{1-e^{-ru_t}}{1-e^{-r}},
\qquad
u_t=\frac{t-t_{\rm h}}{T-1-t_{\rm h}}.
\label{eq:normalized_schedule}
\end{equation}
The stated domains make every denominator nonzero and give
$\lambda_{t_{\rm h}}=0$ and $\lambda_{T-1}=\lambda_{\max}$.
The exponential form is smooth, monotone, and exactly normalized at both endpoints. The parameter $r$ controls the ramp curvature, and $\lim_{r\to0^+}(1-e^{-ru})/(1-e^{-r})=u$, so the linear ramp is recovered as a limiting case. Positive $r$ places more of the increase near the beginning of the hardening phase and then approaches the selected maximum smoothly. This supplies a one-parameter family of ramp shapes without changing the activation epoch or terminal penalty.

For phase offsets $\phi_i\in\mathbb R$ understood modulo $2\pi$,
\begin{equation}
\mathcal L_{\rm geom}(\bm\theta)
=\frac1N\sum_{i=1}^{N}
\sin^2\!\left[\frac K2(\theta_i-\phi_i)\right].
\label{eq:regularizer}
\end{equation}
Each scalar summand has exactly $K$ zero-valued minima on $S^1$. The full regularizer has the product zero set
\begin{equation}
\begin{aligned}
\mathcal W_{K,i}(\phi_i)
&=\left\{\wrap_{[-\pi,\pi)}\!\left(\phi_i+\frac{2\pi j}{K}\right):\right.\\[-2pt]
&\hspace{27mm}\left.j=0,\ldots,K-1\right\},\\
\mathcal Z_K(\bm\phi)&=\prod_{i=1}^{N}\mathcal W_{K,i}(\phi_i),
\qquad |\mathcal Z_K(\bm\phi)|=K^N.
\end{aligned}
\label{eq:regularizer_minima}
\end{equation}
Moreover,
\begin{equation}
\frac{\partial\mathcal L_{\rm geom}}{\partial\theta_i}
=\frac{K}{2N}\sin[K(\theta_i-\phi_i)],\qquad
\left.
\frac{\partial^2\mathcal L_{\rm geom}}{\partial\theta_i^2}
\right|_{\bm\theta\in\mathcal Z_K}
=\frac{K^2}{2N}.
\label{eq:regularizer_scaling}
\end{equation}
These are minima of $\mathcal L_{\rm geom}$ alone; a finite-weight total loss in Eq. \eqref{eq:tcq_loss} need not attain its minimum on the codebook. At fixed $\lambda_t$, changing $K$ changes not only codebook spacing but also the maximum coordinate-gradient scale and the well curvature. The experiments set $\phi_i=0$, so every coordinate-wise penalty well is aligned with Eq. \eqref{eq:codebook}.

\begin{figure*}[t]
\centering
\includegraphics[width=0.92\textwidth]{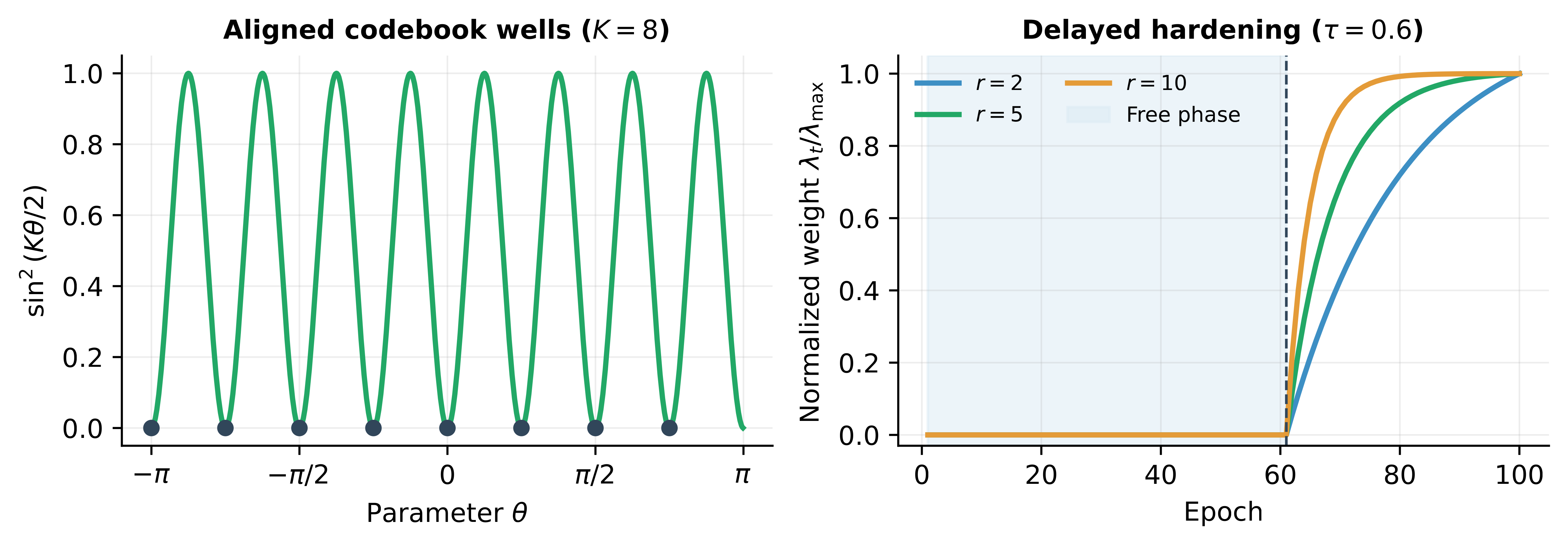}
\caption{T-CQ\ penalty geometry and delayed schedule. Left: the $K$ periodic wells for one coordinate coincide exactly with the deployed angular codewords. Right: the penalty is zero during free optimization; after epoch $t_{\rm h}$, the normalized exponential ramp increases smoothly from zero and reaches $\lambda_{\max}$ at the final epoch.}
\label{fig:method}
\end{figure*}

At each epoch $t\geq t_{\rm h}$, the raw checkpoint is hard-projected and evaluated on validation data. Checkpoints are ordered lexicographically by: (i) highest hard-deployed validation accuracy, (ii) lowest hard-deployed validation MSE, (iii) smallest mean circular distance to the codebook, and (iv) earliest epoch. Python's \texttt{round} is applied to stored floating-point accuracy and MSE at seven decimal places and to distance at 12 decimal places. These resolutions define numerical tie buckets for the recorded protocol and reduce the influence of sub-resolution differences, but an arbitrarily small change can cross a rounding boundary; they do not guarantee identical trajectories or selections across processor architectures and numerical kernels. Remaining ties across otherwise identical schedule candidates are resolved by their declared order in Eq. \eqref{eq:tcq_grid}. Test data are evaluated only after selection.

\subsection{Optional Fixed-Target Circular Relaxation}
Some deployment interfaces may move a continuous control value toward its chosen codeword before the hard snap. For $S\in\mathbb N_0$, $0<\alpha\leq1$, and the exact fixed target $c_i^\star=Q_i(\theta_i^{(0)})$, define
\begin{equation}
\begin{aligned}
a_i^{(s)}&=\wrap_{[-\pi,\pi)}(c_i^\star-\theta_i^{(s)}),\\
\theta_i^{(s+1)}&=\wrap_{[-\pi,\pi)}\!\left(
\theta_i^{(s)}+\alpha a_i^{(s)}\right),
\quad s=0,\ldots,S-1.
\end{aligned}
\label{eq:relaxation}
\end{equation}

\begin{lemma}[Relaxation preserves the exact nearest codeword]
\label{lem:relaxation}
For the uniform codebook and exact quantizer Eq. \eqref{eq:quantizer},
\begin{equation}
\dsone(\theta_i^{(S)},c_i^\star)
=(1-\alpha)^S\dsone(\theta_i^{(0)},c_i^\star),
\end{equation}
and $Q_i(\theta_i^{(S)})=c_i^\star$. At an exact initial Voronoi tie, the lowest-index rule fixes $c_i^\star$, and every $\alpha>0$ moves the first iterate strictly into that codeword's cell.
\end{lemma}

\begin{proof}
Let $e_s=\wrap_{[-\pi,\pi)}(\theta_i^{(s)}-c_i^\star)$. Exact nearest projection gives $|e_0|\leq\pi/K$. The update follows the selected shortest arc and satisfies $e_{s+1}=(1-\alpha)e_s$ without leaving the selected closed Voronoi cell; after one positive step an initial boundary point is in its interior. Iteration proves both claims.
\end{proof}

For $\alpha=0.5$ and $S=3$, the residual circular distance is exactly $1/8$ of its initial value. Because the target is fixed, this relaxation cannot change the exact final hard codeword and is not a source of the discrete-model improvements reported below. The reported experiments do not apply this transition: they export the validation-selected checkpoint directly through $\widehat Q$. The relaxation is included only for deployment interfaces that require intermediate settings before the final hard codeword.

\subsection{Training and Deployment Procedure}
The T-CQ\ procedure is:
\begin{enumerate}
\item Initialize the circuit parameters, optimizer, and phase-zero $K$-level codebook.
\item For $t=0,\ldots,T-1$, compute $\lambda_t$ from Eq. \eqref{eq:normalized_schedule}, update the parameters using Eq. \eqref{eq:tcq_loss}, and, when $t\geq t_{\rm h}$, project the raw checkpoint with $\widehat Q$ and apply the declared validation ordering.
\item Retain the validation-selected raw checkpoint. An interface requiring intermediate settings may optionally apply Eq. \eqref{eq:relaxation}; this step is omitted in the reported experiments.
\item Export the deterministic hard circuit $\widehat Q(\bm\theta)$ and evaluate the held-out test set once.
\end{enumerate}

\section{Experimental Design}\label{setup}

\subsection{Circuit and Learning Task}
All classification experiments use four qubits and two variational layers. Inputs are encoded by $R_y(x_i)$ on the available feature wires. Each layer applies
\begin{equation}
V_\ell(\bm\theta)=\bigotimes_{q=0}^{3}
R_z(\theta_{\ell,q,2})R_y(\theta_{\ell,q,1}),
\end{equation}
followed by the sequential CNOT ring
$0\!\rightarrow\!1\!\rightarrow\!2\!\rightarrow\!3\!\rightarrow\!0$.
The model score is $s_{\bm\theta}(x)=\langle Z_0\rangle$ and labels are encoded as $\{-1,+1\}$. Prediction uses $\widehat y_{\bm\theta}(x)=\operatorname{sgn}_{+}(s_{\bm\theta}(x))$, where $\operatorname{sgn}_{+}(z)=+1$ for $z\geq0$ and $-1$ otherwise; in particular, $\operatorname{sgn}_{+}(0)=+1$. The task loss is full-batch MSE. Each model has $N=2\times4\times2=16$ trainable angles, initialized independently from $\operatorname{Unif}[-\pi,\pi)$ under the matched seed, and is optimized for 100 epochs with Adam at learning rate $0.05$.

\begin{figure*}[t]
\centering
\includegraphics[width=0.92\textwidth]{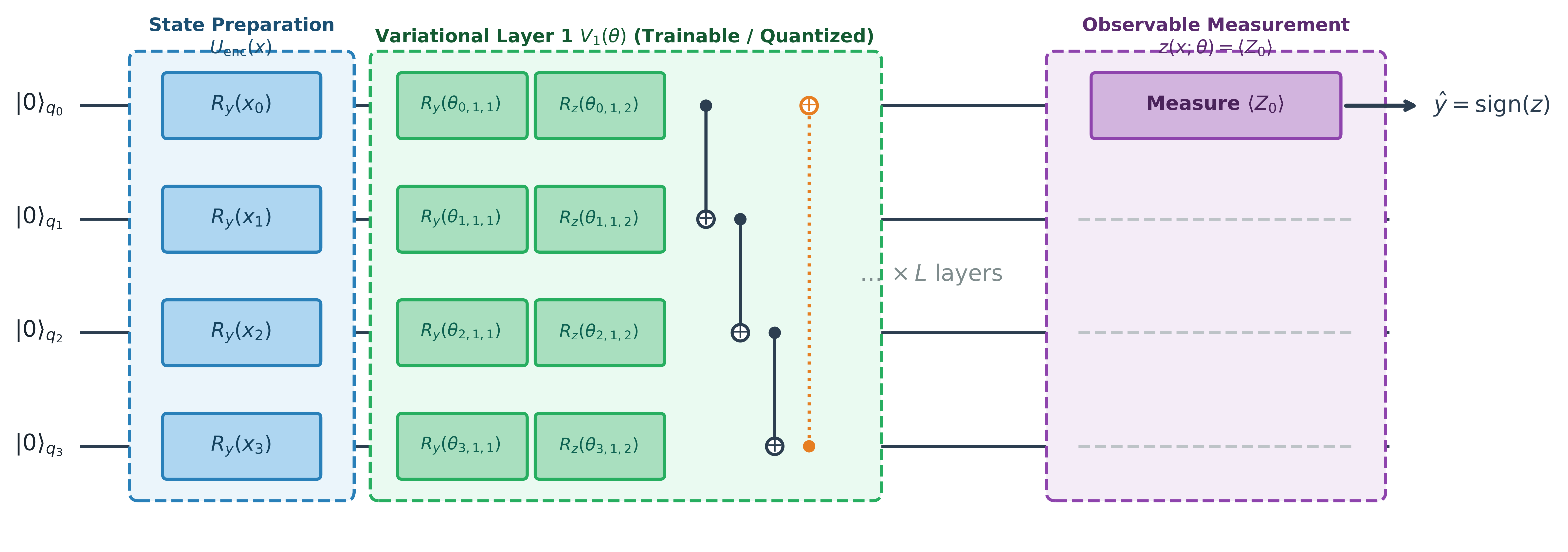}
\caption{Four-qubit classifier used in every experiment: explicit $R_y$ angle embedding, two $R_y$-$R_z$ variational layers with a CNOT ring, and the $Z_0$ score.}
\label{fig:circuit}
\end{figure*}

\subsection{Data and Preprocessing}
The ten matched seeds are
\begin{equation*}
\{42,123,7,99,2024,31415,2718,1618,808,1337\}.
\end{equation*}
Every seed uses a stratified 60\%/15\%/25\% train/validation/test split.
\begin{itemize}
\item \textbf{Two Moons:} 200 samples with noise standard deviation $0.15$; 120/30/50 samples are used for train/validation/test.
\item \textbf{Iris:} the first two classes of Fisher's Iris data \cite{fisher1936}, giving 100 samples and a 60/15/25 split.
\item \textbf{Sampled MNIST 3-vs-6 subset:} OpenML \texttt{mnist\_784}, version 1, supplies the complete 70,000-row source pool \cite{lecun1998}. The original 60,000/10,000 MNIST partition is not preserved. Within each seed, 100 threes and 100 sixes are sampled without replacement before the new stratified 120/30/50 split. Pixel intensities are divided by 255. Four-component PCA with the full-SVD solver is fitted only on the 120 training images, then applied to validation and test images. Tables use the shorter label ``MNIST 3 vs. 6'' for this sampled subset.
\end{itemize}
For every dataset, a min--max affine transform is fitted on the training fold so that each training-feature minimum and maximum map to $0$ and $\pi$. The same transform is applied to validation and test folds without clipping, so held-out values are not assumed to remain inside that interval. The first split uses random state equal to the seed; the train/validation split uses seed plus 10,000. MNIST sampling, splitting, initialization, and all method comparisons are matched within seed.

\subsection{Methods and Capacity Controls}
The primary comparison contains:
\begin{enumerate}
\item \textbf{Continuous:} unconstrained training; the checkpoint is selected by validation accuracy, validation MSE, and epoch.
\item \textbf{Circular-PTQ:} coordinate-wise projection of the continuous checkpoint onto Eq. \eqref{eq:codebook} using Eq. \eqref{eq:circular_distance}, with no retraining.
\item \textbf{Prior-Fitted Coordinate Circular $k$-Means (Prior-Fitted CKM):} Let $M_{\rm p}=500$ and draw $\bm z^{(m)}\sim\operatorname{Unif}([ -\pi,\pi)^{16})$ independently of all task-trained parameters. For each coordinate $i$, ten restarts are initialized by selecting $K$ distinct values from $\{z_i^{(m)}\}_{m=1}^{M_{\rm p}}$. Each iteration assigns sample $m$ to the smallest-index center minimizing $\dsone^2(z_i^{(m)},c_{i,j})$. A nonempty cluster center is updated to
\begin{equation*}
c_{i,j}\leftarrow
\operatorname{atan2}\!\left(
\sum_{m:a_i^{(m)}=j}\sin z_i^{(m)},
\sum_{m:a_i^{(m)}=j}\cos z_i^{(m)}
\right).
\end{equation*}
An empty cluster retains its preceding center, and a zero resultant uses $\operatorname{atan2}(0,0)=0$. Assignment by $\dsone^2$ is also assignment by squared chordal distance because both are monotone in angular separation on $[0,\pi]$; the circular mean is the corresponding chordal-centroid update. Iteration stops after 200 updates or when the largest circular center displacement is below $10^{-7}$. The retained restart has the smallest final geodesic inertia $\sum_m\min_j\dsone^2(z_i^{(m)},c_{i,j})$, with the earliest restart retained on an exact tie. The continuous checkpoint is then projected coordinate-wise onto these ordered centers using Eq. \eqref{eq:numerical_quantizer}. This control is matched to the uniform grid in per-coordinate cardinality---both expose $K$ values per coordinate and $K^{16}$ product combinations---but not in storage because it retains $16K$ learned real centroids. It is independent of the task-trained checkpoint and therefore not task adaptive. The baseline seed is the matched seed for $K=8$ and seed $+100000K$ for $K\in\{4,16\}$. In every case the prior stream is advanced by the archived $16\times K$ random-anchor draw before sampling the 500 prior vectors, and coordinate $i$ uses the baseline seed plus $i$ for restart initialization.
\item \textbf{STE-QAT:} the training forward circuit uses the exact hard uniform circular quantizer at every step, while backpropagation replaces its derivative by the identity \cite{bengio2013}. The latent vector is optimized by the same Adam configuration, and the numerical-map validation checkpoint is selected across all 100 epochs by the declared rule. The reported $\gamma_{\rm map}$ is the latent-to-export separation defined in Eq. \eqref{eq:gamma_map}.
\item \textbf{T-CQ:} training from the same initialization as Continuous, using Eq. \eqref{eq:tcq_loss} and hard deployment onto the same phase-zero uniform codebook as Circular-PTQ.
\end{enumerate}
Continuous, STE-QAT, and every T-CQ\ candidate start from the same circuit initialization within a dataset/seed/$K$ condition. Their task-optimization trajectories use the same split, full-batch MSE, Adam configuration, learning rate, and 100-epoch \emph{per-run} horizon. Circular-PTQ and Prior-Fitted CKM instead transform the selected Continuous checkpoint and perform no separate classifier-training trajectory. The comparison is not matched in total tuning computation: T-CQ\ trains all 11 schedule candidates, totaling 1,100 optimizer epochs per condition, whereas STE-QAT trains one 100-epoch trajectory and the Continuous trajectory is reused by the post-training controls. The results therefore evaluate the declared validation-selected T-CQ\ protocol, not equal-compute optimizer efficiency. Its candidate set is
\begin{multline}
\{(0.4,5,0.1),(0.4,5,0.3),(0.4,5,0.6),(0.6,2,0.1),\\
(0.6,5,0.1),(0.6,5,0.3),(0.6,5,0.6),(0.6,10,0.3),\\
(0.8,5,0.1),(0.8,5,0.3),(0.8,5,0.6)\}.
\label{eq:tcq_grid}
\end{multline}

The candidate set was determined prior to testing of performance and is limited to a finite sensitivity grid (i.e., it is not a complete search nor a general default). The deployment interface will initially determine K for each new task. The activation fraction $\tau$ determines how long the task-only warm up period lasts; r specifies the shape of the ramp during the activation phase; and $\gamma_{max}$ specifies the maximum value of the final regularization strength. As these parameters are only valid when validated with a specified candidate set and the defined tie breaking method; they cannot be changed at the time of testing. Since K affects both the scaling of the regularizer's gradient and the curvature of its wells, this fixed $\gamma_{max}$ grid examines the geometry of the penalty as well as the resolution simultaneously. This grid does not examine a causal relationship of K alone. An examination that isolates the resolution in relation to K requires an explicit normalization on a penalty sweep, which is beyond what can be examined within this controlled study.


\subsection{Reporting and Verification}
Tables and error bars report mean $\pm$ sample standard deviation over ten seeds. For a paired comparison, let $d_s$ be the within-seed difference, with sample mean $\bar d$ and sample standard deviation $s_d$. The reported nominal interval is
\begin{equation}
\bar d\ \pm\ t_{0.975,9}\frac{s_d}{\sqrt{10}}.
\label{eq:paired_t_interval}
\end{equation}
Exact 95\% repeated-sampling coverage requires independent, identically distributed Gaussian paired differences; approximate coverage ordinarily relies on corresponding independence and normal-approximation assumptions. Those conditions are not established here: the ten seeds are fixed protocol instances and, especially for Iris, splits reuse and overlap observations from the same finite source data. The intervals are therefore descriptive summaries of these ten paired runs, not population-wide confidence guarantees. No multiplicity adjustment is applied.

All expectations over $x$ are empirical means over the held-out test fold, used only after validation selection. The audit recomputed every aggregate and checked split sizes, seeds, selection, accuracy granularity, all three Fubini--Study quantities, the infidelity equivalence Eq. \eqref{eq:infidelity_equivalence}, the Minkowski upper bound Eq. \eqref{eq:gamma_decomposition}, and the observable-MSE bound Eq. \eqref{eq:mse_bound}.

The primary numerical record is the archived noiseless four-qubit state-vector execution on an ARM64 CPU using Python 3.14.3, PyTorch 2.12.0, NumPy 2.4.6, scikit-learn 1.8.0, four PyTorch threads, single-precision real parameters, and complex single-precision state vectors. It required 925.4 seconds and reached 1.08 GiB peak process resident memory. All primary tables and figures are derived exclusively from this record.

We additionally performed a complete environmental reexecution using the same seeds, data and split rules, schedule grid, optimizer, selection rule, and test isolation on an x86-64 Windows CPU with Python 3.14.6, PyTorch 2.12.0+cpu, NumPy 2.4.6, and scikit-learn 1.8.0. That execution used 16 PyTorch threads and is analyzed separately rather than pooled with the primary run. Because processor architecture, Python patch level, PyTorch build, and thread count changed jointly, this comparison assesses portability of the empirical conclusion rather than the causal effect of any one environment variable.
\section{Results}\label{results}
Each reported predictions and state distances are based on noiseless state-vector simulation and exact expectation values.
\begin{table*}[t]
\caption{Primary-execution ten-seed test results at $K=8$. Values are mean $\pm$ sample SD. $\gamma_{\rm map}$ is the raw/latent-to-export projection diagnostic and $\gamma_{\rm ref}$ uses the independently selected continuous reference. For Continuous, Circular-PTQ, Prior-Fitted CKM, and T-CQ, $\gamma_{\rm map}=\gamma_{\rm loc}$; for STE-QAT it compares latent parameters with the exported hard circuit.}
\label{tab:main_k8}
\centering
\footnotesize
\begin{tabular}{llccc}
\toprule
Dataset & Method & Test accuracy & $\gamma_{\rm map}$ & $\gamma_{\rm ref}$ \\
\midrule
Two Moons & Continuous & $0.862\pm0.054$ & $0.0000\pm0.0000$ & $0.0000\pm0.0000$ \\
 & Circular-PTQ & $0.826\pm0.045$ & $0.1428\pm0.0501$ & $0.1428\pm0.0501$ \\
 & Prior-Fitted Coord. CKM & $0.862\pm0.035$ & $0.1741\pm0.0326$ & $0.1741\pm0.0326$ \\
 & STE-QAT & $0.798\pm0.092$ & $0.1978\pm0.0495$ & $1.2233\pm0.5766$ \\
 & T-CQ & $0.820\pm0.044$ & $0.0110\pm0.0213$ & $0.3233\pm0.3356$ \\
\midrule
Iris & Continuous & $0.992\pm0.017$ & $0.0000\pm0.0000$ & $0.0000\pm0.0000$ \\
 & Circular-PTQ & $0.944\pm0.060$ & $0.1281\pm0.0533$ & $0.1281\pm0.0533$ \\
 & Prior-Fitted Coord. CKM & $0.952\pm0.049$ & $0.1933\pm0.0452$ & $0.1933\pm0.0452$ \\
 & STE-QAT & $0.952\pm0.037$ & $0.1965\pm0.0603$ & $1.0647\pm0.5360$ \\
 & T-CQ & $0.960\pm0.038$ & $0.0294\pm0.0518$ & $0.2276\pm0.2099$ \\
\midrule
MNIST 3 vs. 6 & Continuous & $0.972\pm0.023$ & $0.0000\pm0.0000$ & $0.0000\pm0.0000$ \\
 & Circular-PTQ & $0.952\pm0.025$ & $0.1153\pm0.0444$ & $0.1153\pm0.0444$ \\
 & Prior-Fitted Coord. CKM & $0.924\pm0.040$ & $0.1808\pm0.0493$ & $0.1808\pm0.0493$ \\
 & STE-QAT & $0.952\pm0.033$ & $0.1894\pm0.0284$ & $0.9843\pm0.4388$ \\
 & T-CQ & $0.956\pm0.025$ & $0.0104\pm0.0134$ & $0.1792\pm0.1031$ \\
\bottomrule
\end{tabular}
\end{table*}

\begin{table*}[t]
\caption{Primary-execution paired ten-seed differences at $K=8$ with nominal two-sided 95\% $t$ intervals (9 d.f.). All differences are T-CQ minus baseline. For STE-QAT, $\gamma_{\rm map}$ is the latent-to-export diagnostic.}
\label{tab:paired_k8}
\centering
\footnotesize
\begin{tabular}{llcccc}
\toprule
Dataset & Baseline & Accuracy difference & Accuracy W/T/L & $\gamma_{\rm map}$ difference & $\gamma_{\rm ref}$ difference \\
\midrule
Two Moons & Circular-PTQ & $-0.006\,[-0.016,+0.004]$ & 0/8/2 & $-0.1318\,[-0.1754,-0.0881]$ & $+0.181\,[-0.061,+0.422]$ \\
 & STE-QAT & $+0.022\,[-0.030,+0.074]$ & 5/0/5 & $-0.1868\,[-0.2284,-0.1452]$ & $-0.900\,[-1.452,-0.348]$ \\
\midrule
Iris & Circular-PTQ & $+0.016\,[-0.031,+0.063]$ & 1/8/1 & $-0.0986\,[-0.1402,-0.0571]$ & $+0.100\,[-0.051,+0.250]$ \\
 & STE-QAT & $+0.008\,[-0.010,+0.026]$ & 3/6/1 & $-0.1670\,[-0.2326,-0.1015]$ & $-0.837\,[-1.192,-0.482]$ \\
\midrule
MNIST 3 vs. 6 & Circular-PTQ & $+0.004\,[-0.002,+0.010]$ & 2/8/0 & $-0.1049\,[-0.1393,-0.0705]$ & $+0.064\,[-0.002,+0.130]$ \\
 & STE-QAT & $+0.004\,[-0.011,+0.019]$ & 2/6/2 & $-0.1790\,[-0.1996,-0.1584]$ & $-0.805\,[-1.108,-0.502]$ \\
\bottomrule
\end{tabular}
\end{table*}

\begin{figure*}[t]
\centering
\includegraphics[width=0.98\textwidth]{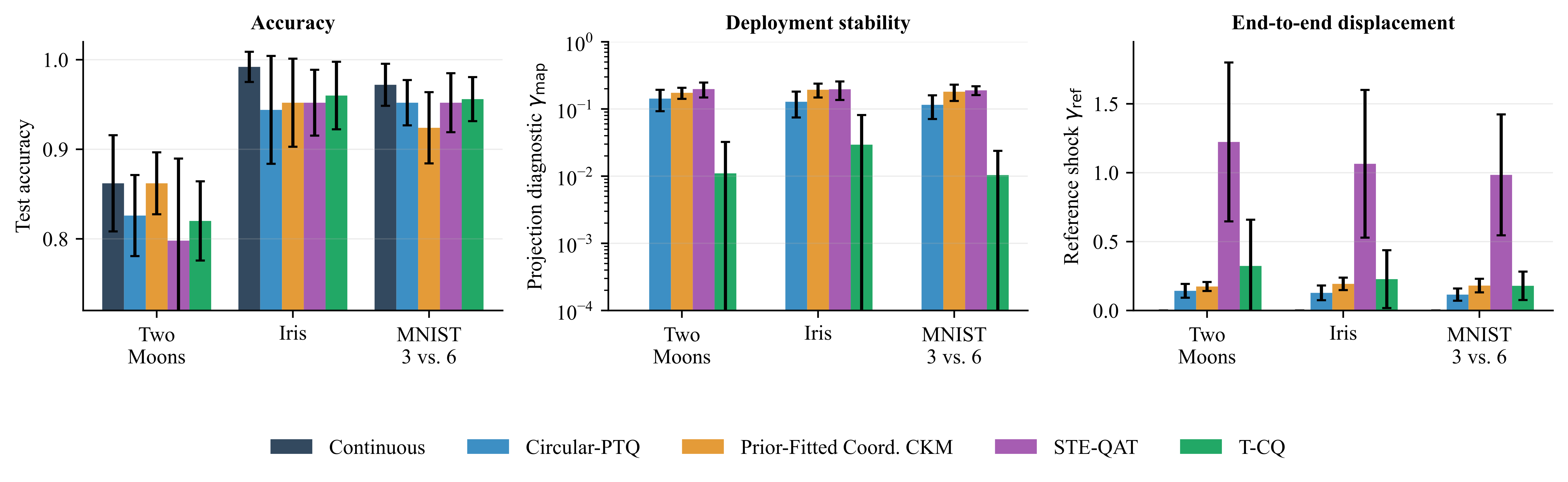}
\caption{Primary-execution ten-seed $K=8$ means with sample-SD error bars. The middle panel is $\gamma_{\rm map}$: it equals local Quantum Shock except for STE-QAT, where it is a latent-to-export diagnostic. The right panel uses the independently selected continuous reference.}
\label{fig:main_k8}
\end{figure*}

\subsection{\texorpdfstring{$K=8$}{K=8}: Matched Baseline Comparison
} 
To summarize the predictions of the T-CQ\ algorithm, we separate prediction accuracy, raw/latent projections, and continuous reference displacement in the primary experiment (see Table~\ref{tab:main_k8}). On the basis of the two Moon dataset, T-CQ\ achieves test accuracies of $0.820\pm0.044$, $0.960\pm0.038$, and $0.956\pm0.025$ for the Iris, MNIST 3-vs-6 datasets. STE-QAT achieves test accuracies of $0.798\pm0.092$, $0.952\pm0.037$, and $0.952\pm0.033$ for the same datasets. The nominal paired T-CQ-minus-STE intervals for accuracy are $+0.022[-0.030,0.074]$, $+0.008[-0.010,0.026]$, and $+0.004[-0.011,0.019]$. 


The nominal paired T-CQ-minus-STE-QAT accuracy interval includes zero for each of the three tasks. Similarly, the T-CQ-minus-Circular PTQ intervals include zero for each of the three tasks. This summary of results from the 10-subject protocol does not show that one approach is better than the other in terms of accuracy; rather they provide an indication of how well each method performed on average across all three task types.

Prior-fitted CKM yields test accuracies of $0.862$, $0.952$, and $0.924$; its nominal paired interval favors CKM on Two Moons, includes zero on Iris, and favors T-CQ\ on MNIST. The task independent learned prior did not provide a consistent ordering based on accuracy.

On average the local Quantum shock values were $0.0110$, $0.0294$, and $0.0104$ for T-CQ,\ while Circular PTQ averaged $0.1428$, $0.1281$, and $0.1153$ for the same data sets (Two Moons, Iris, and MNIST). The ratios of unrounded averages are thus $13.0\times$, $4.4\times$, and $11.1\times$. The nominal paired T-CQ-minus-Circular PTQ mean intervals were negative for all three tasks. In addition, T-CQ\ was less seedwise expensive in terms of local quantum shock in 10/10, 9/10, and 10/10 comparisons on the primary run for the same data sets. These numbers and ratios are descriptive of this particular run and do not represent invariant platform properties.

These comparisons are method level rather than equal total search/compute time comparisons. Both candidates have the same 100 epoch time horizon, however T-CQ\ may select among 11 different validation-evaluated schedules, STE-QAT uses one schedule (trajectory), and the post-training controls reuse the Continuous checkpoint. No claim of runtime efficiency is made.

Additionally, the complete x86-64 re-execution had lower mean T-CQ\ local shock on every task, with lower seed-wise shock in 9/10, 10/10, and 10/10 comparisons respectively. As shown in Table~\ref{tab:environment_k8}, the two executions place side-by-side without pooling. All local and continuous reference paired intervals remain below zero but local-shock ratio changed to $4.9\times$, $5.0\times$, and $7.8\times$. Therefore, while the overall direction remains constant, magnitudes and some individual selections are execution environment dependent.

\begin{table*}[t]
\caption{Environmental reexecution at $K=8$. Local shock values are mean $\pm$ sample SD over the same ten seeds. The ratio is the Circular-PTQ mean divided by the T-CQ mean; intervals are nominal 95\% paired $t$ intervals for T-CQ minus the named baseline. Runs are not pooled.}
\label{tab:environment_k8}
\centering
\scriptsize
\begin{tabular}{lllcccc}
\toprule
Environment & Dataset & Circular-PTQ & T-CQ & Ratio & Local vs. Circular & $\gamma_{\rm ref}$ vs. STE \\
\midrule
ARM64 primary & Two Moons & $0.1428\pm0.0501$ & $0.0110\pm0.0213$ & $13.0\times$ & $[-0.1754,-0.0881]$ & $[-1.452,-0.348]$ \\
 & Iris & $0.1281\pm0.0533$ & $0.0294\pm0.0518$ & $4.4\times$ & $[-0.1402,-0.0571]$ & $[-1.192,-0.482]$ \\
 & MNIST 3 vs. 6 & $0.1153\pm0.0444$ & $0.0104\pm0.0134$ & $11.1\times$ & $[-0.1393,-0.0705]$ & $[-1.108,-0.502]$ \\
\midrule
x86-64 reexecution & Two Moons & $0.1036\pm0.0507$ & $0.0211\pm0.0335$ & $4.9\times$ & $[-0.1260,-0.0390]$ & $[-1.363,-0.449]$ \\
 & Iris & $0.1451\pm0.0505$ & $0.0292\pm0.0510$ & $5.0\times$ & $[-0.1548,-0.0770]$ & $[-0.850,-0.252]$ \\
 & MNIST 3 vs. 6 & $0.1282\pm0.0489$ & $0.0164\pm0.0206$ & $7.8\times$ & $[-0.1483,-0.0754]$ & $[-1.177,-0.366]$ \\
\bottomrule
\end{tabular}
\end{table*}

STE-QAT's displayed $\gamma_{\rm map}$ values describe how far away the latent vector is from the exported hard circuit; they are not considered as measured at runtime.

For the common $\gammaref$ comparison, every nominal T-CQ-minus-STE interval in Table~\ref{tab:paired_k8} is below zero. Therefore, T-CQ\ is generally closer to the independently optimized continuous state than STE-QAT in the primary execution, although it is not universally closer than Circular PTQ.

The environmental re-execution maintains this relative conclusion about reference displacement, as seen in Table~\ref{tab:environment_k8}. T-CQ's accuracies were $0.820$, $0.964$, and $0.956$ while STE-QAT accuracies were $0.802$, $0.956$, and $0.956$ for the same data sets. Thus this implementation provides competitive accuracy and lower reference displacement than STE-QAT in general, but does not achieve superior performance to direct through training.

For T-CQ, mean $\gammadrift$ values were $0.2888$, $0.1697$, and $0.1580$ which were substantially larger than their respective local shocks. This suggests that path movement due to optimization is large in the recorded runs.

Equation~Eq.~(\ref{eq:gamma_decomposition}) represents an upper bound on $\gamma_{\rm drift}$ rather than an additive decomposition and does not indicate any causal contribution.

\subsection{Joint Resolution Sweep and Penalty Geometry Sweep}
\begin{table*}[t]
\caption{Primary-execution joint codebook-resolution and penalty-geometry sweep over ten matched seeds. Values are test mean $\pm$ sample SD.}
\label{tab:resolution}
\centering
\footnotesize
\begin{tabular}{llccc|ccc}
\toprule
& & \multicolumn{3}{c|}{Test accuracy} & \multicolumn{3}{c}{$\gamma_{\rm map}$} \\
Dataset & Method & $K=4$ & $K=8$ & $K=16$ & $K=4$ & $K=8$ & $K=16$ \\
\midrule
Two Moons & Circular-PTQ & $0.786\pm0.130$ & $0.826\pm0.045$ & $0.866\pm0.050$ & $0.3817\pm0.1698$ & $0.1428\pm0.0501$ & $0.0331\pm0.0150$ \\
 & STE-QAT & $0.698\pm0.102$ & $0.798\pm0.092$ & $0.840\pm0.066$ & $0.5580\pm0.0981$ & $0.1978\pm0.0495$ & $0.0415\pm0.0116$ \\
 & T-CQ & $0.832\pm0.053$ & $0.820\pm0.044$ & $0.870\pm0.063$ & $0.0251\pm0.0228$ & $0.0110\pm0.0213$ & $0.0053\pm0.0088$ \\
\midrule
Iris & Circular-PTQ & $0.896\pm0.085$ & $0.944\pm0.060$ & $0.976\pm0.028$ & $0.4604\pm0.1232$ & $0.1281\pm0.0533$ & $0.0403\pm0.0105$ \\
 & STE-QAT & $0.852\pm0.108$ & $0.952\pm0.037$ & $0.976\pm0.028$ & $0.6398\pm0.1183$ & $0.1965\pm0.0603$ & $0.0620\pm0.0174$ \\
 & T-CQ & $0.856\pm0.105$ & $0.960\pm0.038$ & $0.976\pm0.028$ & $0.1985\pm0.1661$ & $0.0294\pm0.0518$ & $0.0041\pm0.0062$ \\
\midrule
MNIST 3 vs. 6 & Circular-PTQ & $0.946\pm0.033$ & $0.952\pm0.025$ & $0.966\pm0.021$ & $0.3620\pm0.1027$ & $0.1153\pm0.0444$ & $0.0308\pm0.0139$ \\
 & STE-QAT & $0.838\pm0.184$ & $0.952\pm0.033$ & $0.970\pm0.029$ & $0.7426\pm0.1775$ & $0.1894\pm0.0284$ & $0.0565\pm0.0107$ \\
 & T-CQ & $0.946\pm0.033$ & $0.956\pm0.025$ & $0.968\pm0.033$ & $0.0208\pm0.0160$ & $0.0104\pm0.0134$ & $0.0075\pm0.0147$ \\
\bottomrule
\end{tabular}
\end{table*}

\begin{figure*}[t]
\centering
\includegraphics[width=0.98\textwidth]{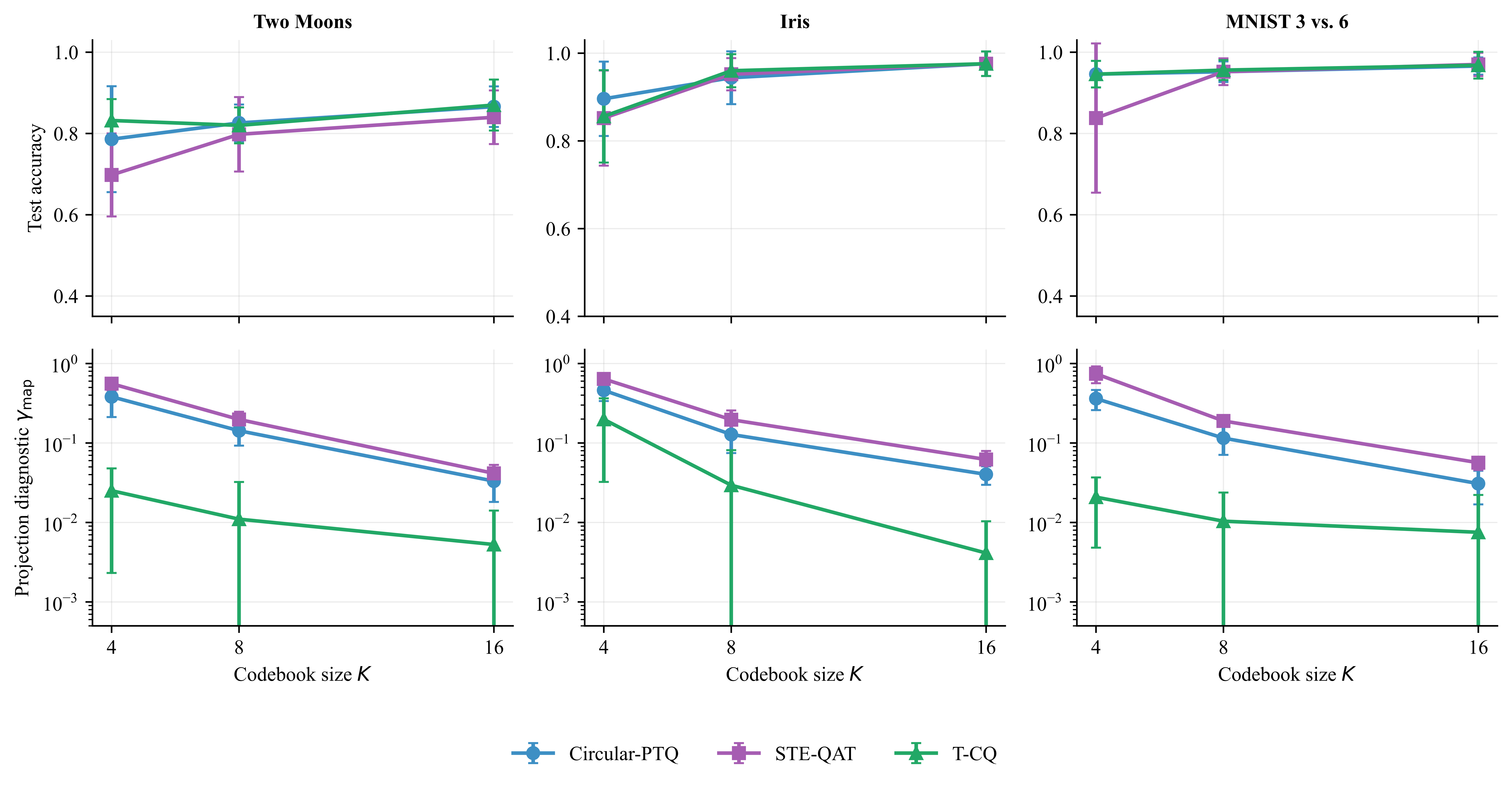}
\caption{primary execution ten seed joint codebook resolution (and) penalty geometry sweep. Sample standard deviation error bars. For STE-QAT, the lower row is the latent to export $\gamma_{\rm map}$ diagnostic. }
\label{fig:resolution}
\end{figure*}

Table~\ref{tab:resolution} and Fig.~\ref{fig:resolution} show how the number of $k$ is changed. When using circular PTQ, changing $k$ changes the size of the uniform projection grid. However when using T-CQ, it can be seen from Eq. \eqref{eq:regularizer_scaling} that changing $k$ will increase the scaling of the regularizer’s coordinate gradient as $k$ and the scaling of its well curvature as $K^2$; since the $\lambda_{\max}$ candidate grid was not scaled to eliminate this effect, the experiment is a combined resolution/penalty geometry sweep, not a completely causal ablation of angular resolution.

When using T-CQ, STE QAT, or circular PTQ at $k = 4,$ we see the following accuracies: $0.832 / 0.698 / 0.786$ on Two moons; $0.856 / 0.852 / 0.896$ on Iris; and $0.946 / 0.838 / 0.946$ on mnist. This is a coarse grid regime for STE QAT, and in this regime T-CQ does not consistently outperform circular PTQ. At $k = 16,$ we see the following triples: $0.870 / 0.840 / 0.866$ on Two moons; $0.976 / 0.976 / 0.976$ on Iris; and $0.968 / 0.970 / 0.966$ on mnist. As such, the difference in accuracy between T-CQ, STE QAT, and circular PTQ has largely disappeared. On average across all nine dataset/k regimes, T-CQ had smaller raw to hard shock than circular PTQ and smaller latent to export $\gamma_{\rm map}$ than STE QAT. However the order of the three methods based on accuracy was still dependent on the data set and k.

These results describe a joint sweep and do not identify what portion of these results is due to either codebook spacing or the penalty geometry being dependent upon k.

\subsection{Convergence and Schedule Sensitivity}
\begin{figure*}[t]
\centering
\includegraphics[width=0.98\textwidth]{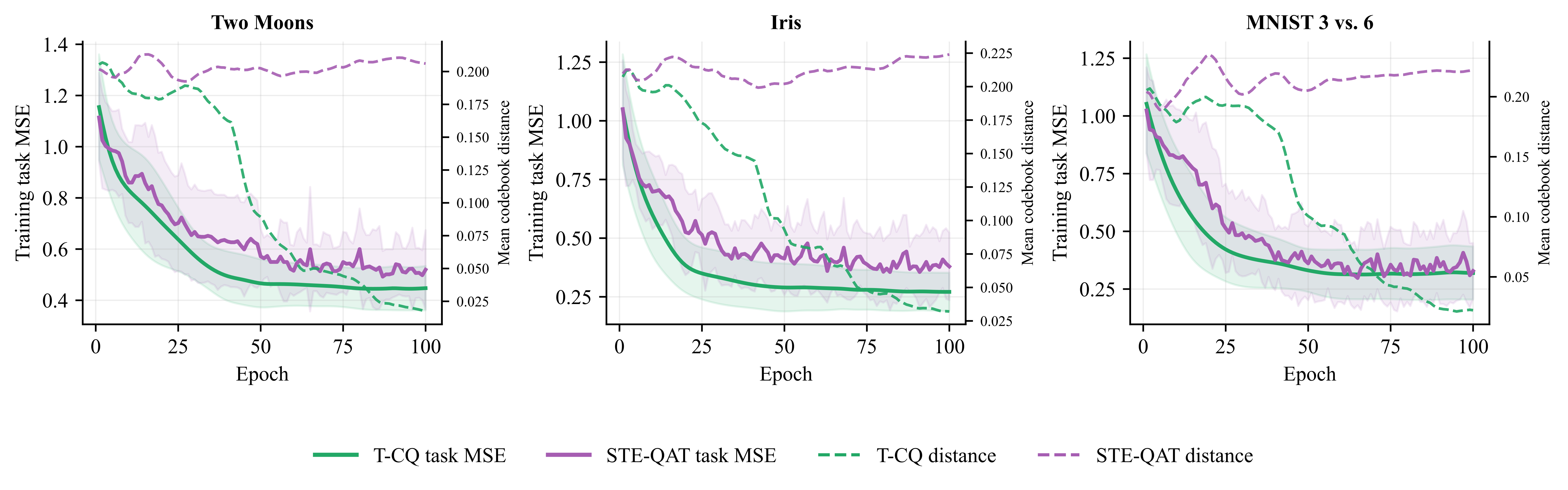}
\caption{$k = 8$ mean training trajectories over ten seeds. Task MSE (with sample SD bands), dashed curves latent/raw distance to hard codebook. T-CQ contracts after delayed hardening while STE-QAT can use a hard forward circuit without moving its latent vector close to the grid.}
\label{fig:convergence}
\end{figure*}

We see from Fig. ~\ref{fig:convergence} that both methods decrease task loss but the methods decrease the loss differently. T-CQ\ first decreases continuously along a task trajectory and then contracts towards the codebook as $\lambda_t$ increases. STE QAT uses hard codewords in every forward pass while allowing its latent vector to remain farther away from them in comparison; thus its latent to export diagnostic may be larger than zero even if the gradient bearing forward model is already hard quantized; as noted after equation (\ref{eq:gamma_map}) numerical near ties prevent an exact zero switching claim for the separate export path.

\begin{table}[t]
\caption{Most frequent validation-selected T-CQ schedule in each ten-seed condition. Counts are out of ten; ties in frequency are resolved only for display by lexicographic order.}
\label{tab:schedule_modes}
\centering
\footnotesize
\begin{tabular}{lccccc}
\toprule
Dataset & $K$ & $\tau$ & $r$ & $\lambda_{\max}$ & Count \\
\midrule
Two Moons & 4 & 0.4 & 5 & 0.6 & 8 \\
Two Moons & 8 & 0.4 & 5 & 0.6 & 8 \\
Two Moons & 16 & 0.4 & 5 & 0.1 & 2 \\
Iris & 4 & 0.4 & 5 & 0.6 & 4 \\
Iris & 8 & 0.4 & 5 & 0.6 & 5 \\
Iris & 16 & 0.4 & 5 & 0.6 & 3 \\
MNIST 3 vs. 6 & 4 & 0.4 & 5 & 0.6 & 6 \\
MNIST 3 vs. 6 & 8 & 0.4 & 5 & 0.6 & 6 \\
MNIST 3 vs. 6 & 16 & 0.4 & 5 & 0.6 & 3 \\
\bottomrule
\end{tabular}
\end{table}

As shown in Table~\ref{tab:schedule_modes} no one schedule outperforms each other on every case. The most common mode is $(0.4,5,0.6)$ for all tasks, selected in $8/10$, $5/10$, and $6/10$ seeds, respectively. Table~\ref{tab:sensitivity_k8} presents the full validation sensitivity results for $k = 8.$ we observe several schedules produce almost indistinguishable validation accuracy but considerably different code book distances, providing further support for selecting jointly validated schedules instead of a universally fixed penalty.

\subsection{Numerical Consistency}
To ensure the quality of our experimental results we performed an extensive set of numerical sanity checks. Our automated auditing tool ran 7,532 finite numerical checks. It evaluated 1,350 observable-mse inequality constraints, 1,350 equivalence constraints for infidelity and 450 Minkowski upper bounds for the 1,350 recorded data points for the combination of dataset, K, seed and method. Furthermore, our tool verified that 105 previously recorded data points at K = 8 were numerically reproducible up to a precision of $2 \times 10^{-6}$; we recalculated the state metrics of T-CQ\ where we had identified a different checkpoint in our seven-decimal validation ordering.

\section{Discussion}\label{discussion}

\subsection{Evidence Summary}
The T-CQ protocol resulted in less than or equal to the amount of mean local deployment shock as did the Circular-PTQ protocol when comparing all 9 data set/code book size combinations listed on table \ref{tab:resolution}. Additionally, the separate x86-64 results shown on table \ref{tab:environment_k8} demonstrate the same mean-shock orderings for the three task executions at k=8 (i.e., T-CQ $<$ PTQ) but have significantly different magnitudes. The test-accuracy comparisons were also somewhat inconsistent and demonstrated that there is no causal relationship between lower local shock and higher accuracy. Overall, the results support an alternative (more narrow) conclusion: for these specific tasks and their respective parameters (circuit, code books, seed values, etc.) delayed hardening resulted in a lower average displacement of the last deterministic projection.

STE-QAT establishes a more precise interpretation. As previously mentioned, STE-QAT uses hard codewords during the forward pass where gradients are being computed and its identity surrogate maintains a large latent-to-grid separation, and, in this particular case, a larger displacement from the continuous reference that was independently trained. On the other hand, T-CQ\ uses a smooth continuous forward model throughout training and then contracts that model toward the grid later on. Thus, there are differences in optimization methods and T-CQ\ had a larger schedule selection budget; none of this indicates that every STE variant or an equally computationally expensive comparison will have identical behavior.

The distinctions among $\gammaloc$, $\gammadrift$, and $\gammaref$ remain important. A small value of $\gammaloc$ does not indicate that T-CQ\ replicated the continuous optimizer: the non-zero $\gammadrift$ indicates that the use of hard codewords resulted in a change to the optimization path. Additionally, corollary~\ref{cor:margin} explains why reduced state displacement provided stability, but did not provide an accuracy guarantee. Also, corollary~\ref{cor:shots} separates this deployment benefit from the additional uncertainty due to the finite number of samples obtained from measuring the system.

\subsection{Why Does the Geometry Matter?}
Circular-PTQ specifically identifies post-training projection itself by mapping the continuous checkpoint selected at test time to the phase zero uniform codebook using circular distance. Prior-Fitted CKM differs from PTQ in terms of how it changes the centroid coordinates of the codebook based on an independent uniform parameter prior, however, it trains the classifier toward those centroids, which makes it task dependent. T-CQ\ introduces yet another mechanism: it modifies the training trajectory such that the raw checkpoint produced at the end of training is close to the uniform codebook. The aligned wells in equation~(\ref{eq:regularizer}) were required to produce this interpretation; if a phase-shifted deployment grid were used, the wells would be aligned with one set of angles, and deployed with another.

Proposition~\ref{prop:infidelity} also places limitations on the novelty claim. Quantum Shock does not define a distinguishability measure for states that is unrelated to fidelity; it is universally equivalent to mean fidelity. The advantage of Quantum Shock here is in providing an organized way to express fidelity: Fubini-Study distance provides a geodesic distance between states, a triangle inequality is used in (\ref{eq:gamma_decomposition}), and a pathway from parameter paths to state displacements. The Bures angle construction provides an extension of the same convention and universal infidelity bounds to density operators; Theorem~\ref{thm:observable} controls observable displacement of density operators through trace distance.

Lemma~\ref{lem:relaxation} also settles what the role of exact circular soft relaxation was. Regardless of whether a fixed target was selected ($S=3$ and $\alpha=0.5$), they reduced residual control distance by $1/8$, but never changed the exact hard codeword, even at exact deterministic ties. Therefore, relaxation is a possible control transition -- not a source of reported discrete-model improvement.

\subsection{Limitations}
This research has been restricted to a very specific case:
one four qubits two layers architecture,
three small binary tasks.
Three different sizes of the codebook, 
ten seeds. 
Test sets have fifty examples (Two Moons) 
and twenty-five samples (MNIST subset)
for each test fold except for Iris where there are twenty-five examples. 
This makes precision of accuracy changes granular. 
These paired intervals represent variations in a given set of seeds under this experimental protocol,
not as part of an arbitrary problem space.

T-CQ\ receives an eleven candidate validation search 
while STE-QAT will receive only one configuration 
of that architecture. 
Therefore, no conclusions regarding superiority can be drawn based on total amount of time spent in tuning. 
Additionally, since changing K sweeps the penalty function's gradient and curvature 
as well as resolution, we cannot draw a direct causal conclusion that K affects performance for T-CQ\ .

All primary simulation runs were done using single-precision arithmetics. 
Because the hard projection operation is a non-differentiable discontinuity at the Voronoi boundary of a codebook partitioning, 
small random differences in an optimization path may lead to a different codeword being selected and thus amplify differences in the shock ratio. 
While the declared bucketing of numerical ties reduces some of these sensitivities, 
it is still possible to get different selections based solely on which floating point environment was used during validation and/or introduce typical uncertainties when selecting a validation set. 
On the other hand, the complete x86-64 execution reproduces the sign of both mean shock effects 
in all nine data-set/resolution combinations, 
the lower mean shock in T-CQ\ , and the same number of decimal points for most primary comparisons. 
However, it did not reproduce every primary count-wise comparison per seed. 
Thus, the direction of shocks is likely much more robust to portability than their magnitudes.

The shock ratios presented above are computed directly from the noiseless simulation of pure-state vectors. 
The Bures angle and finite shot metrics discussed below are analytical generalizations 
of the quantum shock value; they do not constitute empirical claims about a specific noise channel, read-out process or hardware. 
Quantum shock is defined globally and observable agnostically; i.e., if a lower bound on the state displacement implies bounds on every bounded observable through Theorem~\ref{thm:observable}, 
that lower bound on state displacement need not imply identical rankings for all methods when applied to a particular problem. 
Corollary~\ref{cor:shots} discusses only independent projective sampling of a bounded observable 
and does not discuss either biased readout, correlated shots, channel drift or active error-mitigation strategies. 
A similar implementation with a different grid-based converter and update strategy, such as the temperature controlled stochastic low-resolution method proposed by Bhattacharjee et al.~\cite{bhattacharjee2026},
would provide a useful comparative example under a common circular deployment alphabet. 
Device dependent noise characteristics, native control costs, and a full-specified physical-processor-based evaluation are left for future work.
 
\section{Conclusions} \label{conclusion}




We conducted studies on the deterministic finite-resolution deployment of variational circuits using two coupled geometric models: periodic rotation coordinates and the quantum states created in the process. We defined Quantum Shock as the expected squared Fubini-Study distance between the initial state and final state for pure states. We showed that Quantum Shock is upper bounded by a universal multiple of mean infidelity. This provides bounds on both the amount of bounded observable displacement and margin-based prediction stability. Our development of the Bures-angle formulation allows us to extend our distance analysis to include mixed states. We also developed an analytical measurement-stability bound from the finite shot result; however, neither the Bures-angle nor the finite shot extensions were tested in this work using noisy measurements or quantum hardware.

We designed and evaluated T-CQ, which first performs task-only optimization, then introduces a periodic penalty aligned to the task-specific codebooks prior to performing a deterministic circular projection. In the main vector-state study without noise, T-CQ resulted in less mean local deployment shock than Circular PTQ for all three tasks and for each of the three different codebook sizes. When run independently at K=8, x86-64 execution maintained the same relative ordering as the mean shock values; however, the magnitude of these values differed. As stated earlier, test accuracy ranking varied, the search budget per step was larger for T-CQ than it was for STE QAT and K simultaneously affected both the resolution and penalty geometry of the deployments. Therefore, these results demonstrate how delaying "hardening" can be used to decrease localized state displacement under the specific deployment protocol that was tested herein. However, as the validation search budgets differ, these experiments do not establish an establish an equal-compute advantage for T-CQ over STE-QAT.

\begin{table*}[ht]
\caption{K=8 T-CQ sensitivity on validation data. Each cell is the ten-seed mean best hard-deployed validation accuracy / mean circular distance for that fixed schedule after validation-only checkpoint selection. Test data are excluded.}
\label{tab:sensitivity_k8}
\centering
\scriptsize
\begin{tabular}{ccc|cc|cc|cc}
\toprule
$\tau$ & $r$ & $\lambda_{\max}$ & \multicolumn{2}{c|}{Two Moons} & \multicolumn{2}{c|}{Iris} & \multicolumn{2}{c}{MNIST 3 vs. 6} \\
 & & & Acc. & Dist. & Acc. & Dist. & Acc. & Dist. \\
\midrule
0.4 & 5 & 0.1 & 0.837 & 0.0510 & 0.980 & 0.0513 & 0.960 & 0.0494 \\
0.4 & 5 & 0.3 & 0.837 & 0.0340 & 0.953 & 0.0248 & 0.960 & 0.0323 \\
0.4 & 5 & 0.6 & 0.837 & 0.0128 & 0.953 & 0.0273 & 0.957 & 0.0181 \\
0.6 & 2 & 0.1 & 0.830 & 0.0506 & 0.973 & 0.0560 & 0.970 & 0.0451 \\
0.6 & 5 & 0.1 & 0.830 & 0.0514 & 0.973 & 0.0543 & 0.970 & 0.0454 \\
0.6 & 5 & 0.3 & 0.827 & 0.0418 & 0.973 & 0.0285 & 0.953 & 0.0281 \\
0.6 & 5 & 0.6 & 0.827 & 0.0365 & 0.973 & 0.0210 & 0.953 & 0.0174 \\
0.6 & 10 & 0.3 & 0.827 & 0.0419 & 0.973 & 0.0298 & 0.953 & 0.0271 \\
0.8 & 5 & 0.1 & 0.823 & 0.0382 & 0.973 & 0.0667 & 0.963 & 0.0511 \\
0.8 & 5 & 0.3 & 0.823 & 0.0295 & 0.973 & 0.0384 & 0.963 & 0.0360 \\
0.8 & 5 & 0.6 & 0.823 & 0.0237 & 0.973 & 0.0432 & 0.963 & 0.0254 \\
\bottomrule
\end{tabular}
\end{table*}

\FloatBarrier
\clearpage

\appendices
\section{Relaxation and Numerical Audit}\label{relaxation}
The exact statement in Lemma~\ref{lem:relaxation} was checked at $K=8$, $\alpha=0.5$, and $S=3$ on 20,000 random angles and all eight analytical Voronoi midpoints. Separately, the floating-point implementation used the $\varepsilon_{\rm tie}=10^{-12}$ rule in Eq. \eqref{eq:numerical_quantizer}. In this finite test no final numerical codeword changed, and the maximum deviation from the exact residual factor $1/8$ was below $2\times10^{-12}$. This empirical check does not turn the $\varepsilon_{\rm tie}$-nearest map into an exact nearest-neighbor map or extend the lemma to arbitrary approximate-tie inputs.

Every displayed aggregate was recomputed from the ten per-seed records using sample SD. The 7,532-check audit recomputed validation-only selection and accuracy granularity and enumerated the pure-state specialization of the observable-MSE inequality Eq. \eqref{eq:mse_bound}, the infidelity-equivalence bounds Eq. \eqref{eq:infidelity_equivalence}, and the Minkowski upper bound Eq. \eqref{eq:gamma_decomposition}. The mixed-state extension, finite-shot corollary, pointwise and first-moment observable bounds, and margin-distribution corollary were proved analytically but were not separately enumerated among the stored-data checks. Test data were not used for checkpoint or schedule selection.

\begingroup
\renewcommand{\footnotesize}{\fontsize{8}{8.2pt}\selectfont}
\bibliographystyle{IEEEtran}
\bibliography{references}
\endgroup

\end{document}